\documentclass[11pt,preprint]{aastex}
\newcommand{\be}{\begin{equation}}
\newcommand{\ba}{\begin{eqnarray}}
\newcommand{\ee}{\end{equation}}
\newcommand{\ea}{\end{eqnarray}}
\newcommand{\del}{\delta}
\newcommand{\Del}{\Delta}
\newcommand{\etal}{et al.\ }
\newcommand{\etalb}{et al.}

\newcommand\Omm{{\Omega_m}}
\newcommand{\hMpc}{h^{-1}\,{\rm Mpc}}
\newcommand{\hMsun}{h^{-1}\,M_\odot}

\begin{document}
\title {An Analytical Approach to Inhomogeneous Structure Formation}

\author{Evan Scannapieco}
\affil{Department of Astronomy, University of California, Berkeley, CA  94720;
\\{\rm Present address:}\, Osservatorio Astrofisico di Arcetri, Largo E. Fermi,
5, 50125 Firenze, ITALY}

\email{evan@arcetri.astro.it}

\and

\author{Rennan Barkana}
\affil{Canadian Institute for Theoretical Astrophysics, 60 St. George
Street, Toronto, Ontario, M5S 3H8, CANADA; \\{\rm Present address:}\,
School of Physics and Astronomy, Tel Aviv University, Tel Aviv, 
69978, ISRAEL}

\email{barkana@wise.tau.ac.il}

\begin{abstract}

We develop an analytical formalism that is suitable for studying
inhomogeneous structure formation, by studying the joint statistics of
dark matter halos forming at two points. Extending the \citet{bc91}
derivation of the mass function of virialized halos, based on
excursion sets, we derive an approximate analytical expression for the
``bivariate'' mass function of halos forming at two redshifts and
separated by a fixed comoving Lagrangian distance. Our approach also
leads to a self-consistent expression for the nonlinear biasing and
correlation function of halos, generalizing a number of previous
results including those by \citet{kaiser84} and \citet{mo96}. We
compare our approximate solutions to exact numerical results within
the excursion-set framework and find them to be consistent to within
$2\%$ over a wide range of parameters.  Our formalism can be used to
study various feedback effects during galaxy formation analytically,
as well as to simply construct observable quantities dependent on the
spatial distribution of objects.

\end{abstract}

\keywords{cosmology: theory -- galaxies: formation -- 
          large-scale structure of universe -- methods: analytical}

\section{Introduction}

A critical prediction of any theory of structure formation is the mass
function of virialized dark-matter halos.  As the gravitational
collapse of dark matter is thought to be the dominant force in
structure formation, an accurate determination of the number density of
halos as a function of mass and redshift is a critical step towards
understanding the observed
abundances of galaxies, clusters, and other cosmological objects.

In the study of structure formation, two main methods have emerged to
evaluate this quantity: computational methods that solve the equations
of gravitational collapse numerically, and analytical techniques that
approximate these results with simple one-dimensional functions.
While only numerical methods capture the full details of dark matter
collapse, much of our understanding of structure formation relies
instead on analytical techniques.  As such methods are based on simple
assumptions and are easily applied to a large range of models, they
are indispensable both for gaining physical understanding into the
numerical results and exploring the effects of model uncertainties.

The most widely applied method of this type was first developed by
\citet{ps74}. In this model, the abundance of halos at a redshift $z$
is determined from the linear density field by applying a simple model
of spherical collapse to associate peaks in this field with virialized
objects in a full nonlinear treatment.  This simple model, later
refined by \citet{bc91}, \citet{lc93}, and others, has had great
success in describing the formation of structure, reproducing the
numerical results much more accurately than might be expected given
the approximations involved.

Yet this model is intrinsically limited since it can only predict the
average number density of halos, without supplying any information as
to their relative positions. Although this is sufficient for studying
halo evolution, baryonic objects forming within these halos are often
subject to strong environmental effects that are untreatable in this
context. As a simple first-order approximation, many authors have
tried to reconstruct the formation history of baryonic objects by
combining the Press-Schechter approach with average intergalactic
medium (IGM) conditions as a function of redshift,  bathing all
cosmological objects in the same UV background flux or assuming the
same metal pre-enrichment for all galaxies.

Many of the most important environmental effects, however, are in
reality extremely inhomogeneous in nature, being caused by the
nonlinear structures that form within the IGM, and thus primarily
impacting the areas near these structures.  Such interactions between
the IGM and structure formation are often better described as
spatially-dependent feedback loops rather than sudden changes in the
overall average conditions. Processes of this sort include the
formation of the first cosmological objects and the dissociation of
molecular hydrogen \citep[e.g.,][]{hr96,ha00,cf00}, galaxy formation
and photoevaporation during reionization
\citep[e.g.,][]{e92,go97,mr98,bl99} and the impact of galactic outflows
on the formation of neighboring objects \citep[e.g.,][]{st01,sb01}.
While a complete treatment of these issues can only be achieved
numerically, unlike simulation of average quantities, simulations of
structure formation in inhomogeneous environments have no analytical
counterparts with which they can be compared.  This greatly reduces
the parameter space of models that can be studied and leaves us
without a more basic theoretical understanding that can put the
simulation results in a broader context.

Even when objects do not have a large effect on the formation of their
neighbors, issues related to the spatial correlations between halos
often arise when comparing theoretical predictions to the observed
distribution of objects.  Thus analytical estimates of galaxy cluster
correlation functions \citep[e.g.,][]{mjw96}, and of the contribution
of collapsed objects to the angular power-spectrum of the Cosmic
Microwave Background \citep[e.g.,][]{kk99,kn98,sc00}, depend on
supplementing the Press-Schechter number densities with additional
approximate models. While a number of such models exist, along with
accurate fitting formulae for both observational and numerical
correlations \citep[e.g.,][]{e88,ck89,mo96,j98,j99}, all such
techniques represent the grafting of external information onto the
underlying excursion-set calculation.

In this work, we develop an approximate analytical model that can
address these issues. Inspired by the success of the \citet{ps74}
model of structure formation, we return to the linear excursion-set
formalism and consider the collapse of two neighboring points. While
the exact solution to this problem can only be obtained numerically,
we show that such results can be reproduced analytically with great
accuracy by introducing a simple and well-motivated
approximation. These analytical expressions then provide an extension
to the peaks framework that can be used to quickly and easily address
issues of inhomogeneous structure formation.

The structure of this work is as follows.  In \S 2 we review the
derivation of the Press-Schechter formalism used to construct the
average mass function of halos in the universe. In \S 3 we extend this
formalism using an approximate form of the two-point density
distribution, which we compare to exact numerical results.  In \S 4 we
construct the mass function in the neighborhood of an overdense region
with a fixed mass and collapse time, and estimate the nonlinear bias
between halos in this model.  Our conclusions are summarized in \S 5,
and appendixes are included that provide explicit expressions that are
necessary to evaluate our analytical results and describe a publicly
available code that makes it easy to use our formalism for specific
applications. We have added an Erratum (immediately preceding the
appendixes) which adds and discusses a previously missing reference.

\section{Halo Collapse Around a Single Point}

Before addressing the formation of virialized halos at two correlated
points, we first review in this section the approach of \citet{bc91}
which leads to the standard one-point expressions. We work with the
linear overdensity field $\del({\bf x},z) \equiv \rho({\bf
x},z)/\bar\rho(z) - 1$, where ${\bf x}$ is a comoving position in
space, $z$ is the cosmological redshift and $\rho$ is the mass
density, with $\bar \rho$ being the mean mass density. In the linear
regime, the density field maintains its shape in comoving coordinates
and the overdensity simply grows as $\del = \del_i D(z)/D(z_i)$, where
$z_i$ and $\del_i$ are the initial redshift and overdensity, and
$D(z)$ is the linear growth factor [given by eq.~(10) in
\citet{eh99}].  When the overdensity in a given expanding region
becomes non-linear, the expansion halts and the region turns around
and collapses to form a virialized halo.

The time at which the region virializes can be estimated based on the
initial linear overdensity, using as a guide the collapse of a
spherical top-hat perturbation. At the moment at which a top hat
collapses to a point, the overdensity predicted by linear theory is
$\del_{c}=1.686$ \citep{p80} in the Einstein-de Sitter model.  This
value depends weakly on the cosmological parameters, but in the
quantitative plots shown in this paper we fix $\del_{c}=1.686$ for
simplicity. Our analytical expressions, however, do not depend on this
particular choice.

A useful alternative way to view the evolution of density is to
consider the linear density field extrapolated to the present time,
i.e., the initial density field at high redshift extrapolated to the
present by multiplication by the relative growth factor. In this case,
the critical threshold for collapse at redshift $z$ becomes redshift
dependent, \be \del_c(z) = \del_{c} / D(z)\ .\ee We adopt this view,
and throughout this paper the power spectrum $P(k)$ refers to the
initial power spectrum, linearly-extrapolated to the present (i.e.,
not including non-linear evolution).

At a given $z$, we consider the smoothed density in a region around a
fixed point $A$ in space. We begin by averaging over a large mass
scale $M$, or, equivalently, by including only small comoving
wavenumbers $k$. We then lower $M$ until we find the highest value for
which the averaged overdensity is higher than $\del_c(z)$ and assume
that the point $A$ belongs to a halo with a mass $M$ corresponding to
this filter scale.

Note that this description of structure formation is essentially a
Lagrangian one, as it gives us no information as to the motions of
peaks. Instead, this approach provides information only as to the size
of the halo in which the material {\em initially}\, at a point $A$ is
contained. This distinction will prove especially important when
considering two-point quantities, but must be kept in mind even when
interpreting the one-point results.

In this picture
we can derive the mass distribution of halos at a redshift
$z$ by considering the statistics of the smoothed linear density
field.  If the initial density field is a Gaussian random
field and the smoothing is done using sharp $k$-space filters, then the
value of the smoothed $\del$ undergoes a random walk as the cutoff
value of $k$ is increased. If the random walk first hits the collapse
threshold $\del_c(z)$ at $k$, then at a redshift $z$ the point $A$ is
assumed to belong to a halo with a mass corresponding to this value of
$k$. Instead of using $k$ or the halo mass, we adopt as the
independent variable the variance at a particular filter scale $k$,
\be
S_k \equiv \frac{1}{2 \pi^2} \int_0^k dk'\, k'^2\, P(k')\ .
\label{eq:Sk}
\ee

In order to construct the number density of halos in this approach, we
need to find the equation that describes the evolution of the
probability distribution $Q(\del,S_k)$, where $Q(\del,S_k)\, d\del$ is
the probability for a given random walk to be in the interval $\del$
to $\del+d\del$ at $S_k$. Alternatively, $Q(\del,S_k)\, d\del$ can
also be viewed as the trajectory density, i.e., the fraction of the
trajectories that are in the interval $\del$ to $\del+d\del$ at $S_k$,
assuming that we consider a large ensemble of random walks all of
which begin with $\del=0$ at $S_k=0$.

We first examine the evolution of $Q$ in the absence of any
barrier. We consider a small step $\Del S_k$ in $S_k$, during which
$\del$ changes by $\Del \del$. We can obtain $Q(\del,S_k)$ by
integrating over the probability $Q(\del-\Del \del,S_k-\Del S_k)$ that
we started at the point $(\del-\Del \del,S_k-\Del S_k)$, multiplied by
the probability of making the step $(\Del \del, \Del S_k)$ given the
starting point $(\del-\Del \del,S_k-\Del S_k)$. The equation for $Q$
is thus
\be
Q(\del,S_k)= \int_{\Del \del=-\infty}^{\infty} d\, \Del \del\ G(\Del 
\del,\Del S_k)\,
Q(\del-\Del \del,S_k-\Del S_k)\ ,
\label{eq:qd}
\ee
where the probability $G(\Del \del,\Del S_k)$ of making the step does not
depend on the starting point, and is given by the Gaussian
\be
G(\Del \del,\Del S_k) \equiv \frac{1}{\sqrt{2 \pi\, \Del S_k}} \exp
\left[ -\frac{(\Del \del)^2}{2\, \Del S_k} \right]\ .
\ee
To solve the integral equation for $Q$ we expand the term $Q(\del-\Del
\del,S_k-\Del S_k)$ in eq.~(\ref{eq:qd}) with respect to $\del$ and
obtain
\be Q(\del, S_k)=
Q(\del,S_k-\Del S_k) - \frac{\partial Q} {\partial \del} \, \left \langle
\Del \del \right\rangle + \frac{1} {2} \frac{\partial^2 Q} {\partial
\del^2} \, \left\langle(\Del \del)^2 \right\rangle\ ,
\ee
where on the right-hand side all the $Q$'s are evaluated at
$(\del,S_k-\Del S_k)$. The expectation values on the right-hand side
refer to the probability distribution $G(\Del
\del,\Del S_k)$, and they are simply $\left\langle\Del \del \right\rangle=0$ 
and
$\left\langle(\Del \del)^2 \right\rangle=\Del S_k$. The term 
$Q(\del,S_k-\Delta
S_k)$ can then be expanded as $Q(\del,S_k)-\Del S_k \frac{\partial Q} 
{\partial
S_k}$, while we can substitute $S_k$ for $S_k-\Del S_k$ in the other terms 
on
the right-hand side, as this difference would correspond to
higher-order terms and can be neglected.  With these substitutions we
obtain a diffusion equation,
\be
\frac{\partial Q}
{\partial S_k} = \frac{1} {2} \frac{\partial^2 Q} {\partial \del^2},
\label{eq:oneddiff}
\ee
which is satisfied by the usual solution which we label $Q_0$:
\be
Q_0(\del,S_k) = \frac{1}{\sqrt{2 \pi S_k}} \exp \left[
-\frac{\del^2}{2\, S_k} \right]= G(\del,S_k) .
\ee

To determine the probability of halo collapse at a redshift $z$, we
consider the same situation but with an absorbing barrier at
$\del=\nu$, where we set $\nu=\del_c(z)$. The fraction of trajectories
absorbed by the barrier up to $S_k$ corresponds to the total fraction
of mass in halos with masses higher than the value associated with
$S_k$. In this case, the equation satisfied by $Q$ is exactly the same
as in the above derivation, because the chance of being absorbed by
the barrier over the interval $\Del S_k$ goes to zero exponentially as
$\Del S_k \rightarrow 0$, and the barrier has no effect to first
order.  Thus the solution with the barrier in place is given by adding
an extra image-solution:
\be
Q(\nu,\del,S_k)= Q_0(\del,S_k)-Q_0(2 \nu-\del,S_k) .
\label{eq:1}
\ee
Using this expression, we can calculate the fraction of all
trajectories that have passed above the barrier $\nu$ by $S_k$ to be
\be
F(\nu,S_k) = 1-
\int_{-\infty}^{\nu}  d\del\, Q(\nu,\del,S_k) =
2 \int_{\nu}^{\infty} d\del\, Q_0(\del,S_k)\ .
\ee
The differential mass function is then determined by 
\be 
f(\nu,S_k) =
\frac{\partial }{\partial S_k} F(\nu,S_k) = \left(\frac{\partial
Q_0(\del,S_k)}{\partial \del} \right)^{\del=\infty}_{\del=\nu} =
\frac{\nu }{\sqrt{2 \pi} S_k^{3/2}} \exp\left[-\frac{\nu^2}{2 S_k}\right]\
, \label{eq:f1pt} 
\ee 
where we have used the fact that $Q_0$ satisfies
eq.~(\ref{eq:oneddiff}). As $f(\nu,S_k)\, dS_k$ is the probability
that point $A$ is in a halo with mass in the range corresponding to
$S_k$ to $S_k+d S_k$, the halo abundance is then simply
\be
\frac{dn}{dM} = \frac{\bar{\rho}}{M} \left|\frac{d S_k}{d M} \right|
f(\nu,S_k)\ ,
\label{eq:abundance}
\ee
where $dn$ is the comoving number density of halos with masses in the
range $M$ to $M+dM$. The cumulative mass fraction in halos above mass
$M$ is similarly determined to be
\be
\label{eq:PSerfc} F(>M | z)={\rm erfc}\left(\frac{\nu}
{\sqrt{2 S_k}} \right)\ .
\label{eq:fm1point}
\ee

While these expressions were derived in reference to density
perturbations smoothed by a sharp $k$-space filter as given in
eq.~(\ref{eq:Sk}), $S_k$ is often replaced in the final results with
the variance of the mass $M$ enclosed in a spatial sphere of comoving
radius $r$:
\be
\sigma^2(M) = \sigma^2(r) = \frac{1}{2 \pi^2} \int_0^\infty k^2 dk P(k) 
W^2(k r)\ ,
\label{eq:sig}
\ee
where $W(x)$ is the spherical top-hat window function, defined in
Fourier space as
\be
W(x) \equiv 3 \left[ \frac{\sin(x)}{x^3} - \frac{\cos(x)}{x^2} \right]\ .
\label{eq:reW}
\ee 
With this replacement we recover the cumulative mass fraction that
was originally derived in \citet{ps74} simply by
considering the distribution function of overdensities at a single
point, smoothed with a top-hat window function, and integrating from
$\delta_c$ to infinity.  In this derivation the authors were forced to
multiply their result by an arbitrary factor of two, to account for
cases in which collapsed peaks were contained within a collapsed peak
at a larger scale. The excursion-set derivation presented here, based
on \citet{bc91}, properly accounts for such peaks-within-peaks,
however, as well as makes explicit the approximations involved in
working with $\sigma^2(r).$ Strictly speaking, dealing with a
real-space filter requires a complete recalculation of
$f(\nu,\sigma^2)$ which accounts for the correlations intrinsic to
$W(x)$. However, simply replacing $S_k$ with $\sigma^2(r)$ in
eq.~(\ref{eq:abundance}) has been shown to be in reasonable agreement
with numerical simulations \citep[e.g.,][]{katz93}, 
and is thus a standard approximation.

While this standard Press-Schechter mass function, in which $\delta_c
= 1.686$ and $S_k = \sigma^2(r)$, is a useful statistical tool, it is
possible to improve its agreement with simulations by adopting more
complicated choices of these parameters, and adjusting the functional
form of eq.~(\ref{eq:f1pt}).  Finding the ideal model for $\delta_c$,
$S_k$ and $f(\nu,S_k)$ has become somewhat of an art, with many
authors proposing various approaches.  Models have been studied in
which the mass function is modified by fitting to simulations
\citep[]{smt01,jenk01}, incorporating the Zel'dovich approximation
\citep[]{mon95,mon97a,mon97b,ls98,ls99} or even extending the adhesion
approximation \citep[]{men01}, itself an extension to the Zel'dovich
approximation.  While these methods improve the accuracy of the
Press-Schechter technique, they are concerned with single-point
quantities and thus probe an altogether different direction than the
one explored here.  In this investigation, then, we focus on
generalizing the basic excursion-set analysis to the collapse of two
neighboring points; we develop a two-point formalism to which
improvements to the single-point mass function are likely to be
directly applicable.

\section{Two-Point Halo Collapse}

\subsection{Analytic Preliminaries}

\label{sec:sigma8}

\begin{figure}
\plotone{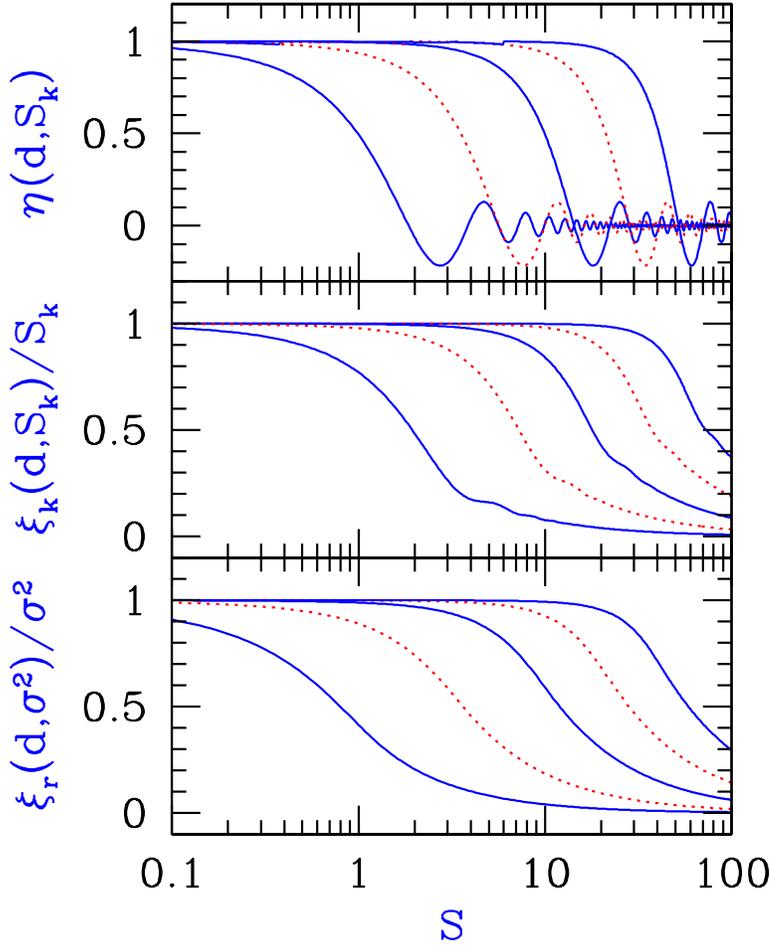}
\caption{
Various correlation functions. {\em Upper panel:}\, each line shows
$\eta(d,S_k)$ as a function of $S_k$ for various values of $d$.  From
left to right $d=10$, 3.3, 1, 0.33, and $0.1\hMpc$ respectively. {\em
Center panel:}\, $\xi_k(d,S_k)/S_k$ as a function of $S_k$ for the
same Lagrangian distances as in the upper panel, again arranged from
left to right. {\em Bottom panel:}\, $\xi_r(d,\sigma^2)/\sigma^2$ as a
function of $\sigma^2$, again for the same distances arranged from
left to right.}
\label{fig:etas}
\end{figure}

Having reviewed the standard derivation of the one-point halo mass
function, we turn to the evolution of two points, separated by a fixed
comoving distance $d$.  Note that this definition of distance is in
Lagrangian space, which is intrinsic to any Press-Schechter type
approach. Thus it is the comoving distance between points $A$ and $B$
at early times, and does not take into account subsequent motions of
these points. In the two-point case, one quantity that enters is the
cross-correlation between two objects identified by sharp $k$-space filters:
\be \xi_k(d,S_k) \equiv \frac{1}{2 \pi^2} \int_0^{k(S_k)} k'^2\, dk'\,
\frac{\sin(k'd)} {k'd}\, P(k')\ ,
\label{eq:xik}
\ee
where the integration limit $k$ is set by $S_k$ through
eq.~(\ref{eq:Sk}). It is also convenient to define
\be
\eta(d,S_k)
\equiv \frac{\sin\left[k(S_k)\, d\right]}{k(S_k)\, d}\ ,
\label{eq:etak}
\ee
so that \be
\xi_k (d,S_k)= \int_{S'=0}^{S_k} \eta(d,S_k')\, dS_k'\ .  \label{eq:xi}
\ee
If we consider a filter $k_1$ at point $A$ and $k_2$ at point $B$, the
cross-correlation involves only those $k$-values common to both
filters, and the result is simply given by eq.~(\ref{eq:xik}) where
the upper integration limit is $k={\rm min}[k_1,k_2]$. Curves of
$\eta(d,S_k)$ and $\xi_k(d,S_k)/S_k$ for various values of $d$ are
shown in the upper two panels of Figure~\ref{fig:etas}. Note that at
small values of $S_k$, when $d \ll k^{-1}(S_k)$, $\eta(d,S_k)$
approaches $1$, and thus $\xi_k$ tends toward $S_k$. At high values of
$S_k$, when $d \gg k^{-1}(S_k)$, the perturbations become
uncorrelated, and thus $\xi_k \ll S_k.$

In this figure and throughout this paper, we illustrate our results
for a cosmological model corresponding to a Cold Dark Matter (CDM)
cosmogony with a non-zero cosmological constant, a choice that is
based mainly on the latest measurements of the Cosmic Microwave
Background \citep[e.g.][]{ba00,net01,pr01}.  We fix $\Omm$ = 0.3,
$\Omega_\Lambda$ = 0.7, $\Omega_b = 0.05$, $\sigma_8 = 0.8$, $h=0.65$,
and $n=1$, where $\Omm$, $\Omega_\Lambda$, and $\Omega_b$ are the
total matter, vacuum, and baryonic densities in units of the critical
density, $\sigma_8 = \sigma (8\, \hMpc)$ as in eq.\ (\ref{eq:sig}),
$h$ is the Hubble constant in units of $100 \, {\rm km} \, {\rm
s}^{-1} \, {\rm Mpc}^{-1}$, and $n$ is the tilt of the primordial
power spectrum, where $n=1$
corresponds to a scale invariant spectrum.  Our results apply quite
generally to any cosmology, however, and thus this model is only an
illustration of our approach.

Just as we needed the real-space variance in the one-point case, many
of the relevant two-point quantities discussed below depend on the
correlation between two spatial filters centered about two points at a
separation $d$. In this case, the standard expression is
\be
\xi_r(d,r_1,r_2) \equiv \frac{1}{2 \pi^2} \int_0^\infty k^2 dk
\frac{\sin(k d)}{k d} P(k) W(k r_1) W(k r_2)\ ,
\label{eq:xi2}
\ee
where $r_1$ and $r_2$ are the radii of the two filters, and $W(x)$ is
again the top-hat window function given by
eq.~(\ref{eq:reW}). Unfortunately, simply substituting this quantity
for $\xi_k$ in analogy with the usual one-point ansatz of substituting
$\sigma^2$ for $S_k$ does not yield an acceptable approximation. 
This is because in the simple limit $d \rightarrow 0$, the correct
one-point results [see eq.~(\ref{eq:merger}) below] are recovered only
if $\xi \rightarrow {\rm min}[\sigma^2(r_1),\sigma^2(r_2)]$. In this
work, then, we instead make use of
\be
\xi_{r_{max}}(d,r_1,r_2)
\equiv \xi_r[d,{\rm max}(r_1,r_2), {\rm max}(r_1,r_2)]\ ,
\label{eq:xirmax}
\ee
which is equal to $\xi_r$ when the two filters have equal radii.

We have also explored an alternative approach, in which we replace
$\xi_r$ with $\xi_k(d,S_k)$ as in eq.~(\ref{eq:xik}), but now choosing
the integration limit $k$ to correspond to real-space filtering, so
that $S_k(k) = \sigma^2(r).$ In this case we define
\be
\xi_{k(r)}(d,r_1,r_2) \equiv
\xi_k\{d,{\rm min}[\sigma^2(r_1),\sigma^2(r_2)] \},
\ee
where $k$ is not proportional to $1/r$ but instead these quantities
are only indirectly related, with $r$ determining $\sigma^2$ which in
turn determines $k$. In this approach, the derivatives of
$\xi_{k(r)}$ with respect to $r_1$ and $r_2$ can be expressed
analytically in terms of $\eta$, reducing the overall computational
load.

These alternative definitions of the correlation function are
illustrated in Figure~\ref{fig:etas} in which we plot
$\xi_r(d,\sigma^2)/\sigma^2 \equiv
\xi_r[d,r(\sigma^2),r(\sigma^2)]/\sigma^2$ and
$\xi_k(d,S_k)/S_k$ as functions of $\sigma^2$ and $S_k$ respectively.
This parameterization allows for a direct comparison between $\xi_{k}$
and $\xi_r$, and shows that while these two correlation functions are
rather similar, $\xi_r/\sigma^2$ is smooth while $\xi_k/S_k$ shows
oscillations at intermediate scales. While these oscillations are
small, we have found that they become more pronounced in some of the
two-point quantities discussed below.  Thus we choose in this paper to
base our ansatz on the smoother quantity $\xi_{r_{max}}$, despite the
fact that its derivative must be calculated numerically.

\begin{figure}
\plotone{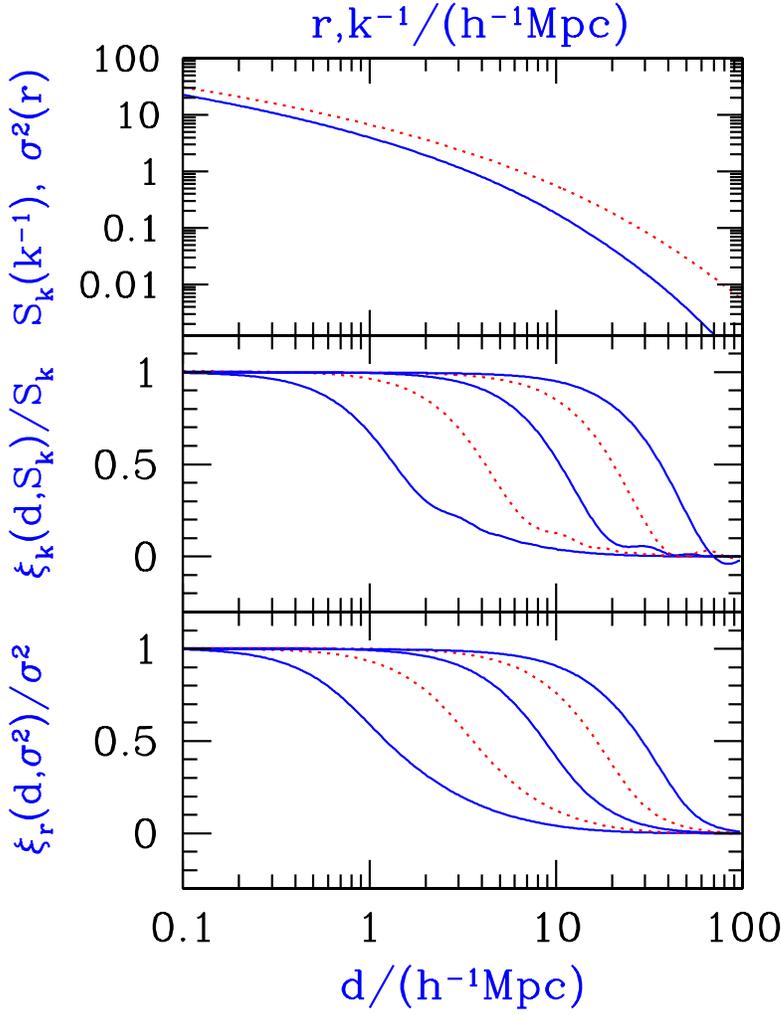}
\caption{
Variance and correlation functions as a function of scale. {\em Upper
panel:}\, The solid and dotted lines show $S_k(k^{-1})$ and
$\sigma^2(r)$ respectively. {\em Center panel:}\, $\xi_k(d,S_k)/S_k$
as a function of $d$ for various values of $S_k$. From left to right
$S_k$ = 10, 3.3, 1.0, 0.33, and 0.1, corresponding to $k^{-1}$ = 0.38,
1.3, 3.4, 6.8, and 12 comoving $\hMpc$ respectively.  {\em Bottom
panel:}\, $\xi_r(d,\sigma^2)/\sigma^2$ as a function of $d$.  From left
to right $\sigma^2$ = 10, 3.3, 1.0, 0.33, and 0.1, corresponding to
$r$ = 0.70, 2.4, 6.5, 14, and 26 $\hMpc$.}
\label{fig:sigs}
\end{figure}

In Fig.\ \ref{fig:sigs} we examine how these quantities vary as a
function of scale.  In the upper panel of this figure, we plot $S_k$
and $\sigma^2$ as functions of $k^{-1}$ and $r$ respectively.  Here we
see that while both functions decrease monotonically with increasing
length scale, $\sigma^2$ is somewhat greater than $S$ at all spatial
values, such that roughly $S_k(1/r) \approx \sigma^2(2 r).$
Thus it is more appropriate to compare $r$ to $2/k$ than $1/k$.  In
the lower two panels of this figure, we plot $\xi_k$ and $\xi_r$ as
functions of Lagrangian distance for various values of $S_k$ and
$\sigma^2$, respectively. Here again we find much the same behavior as
in Figure~\ref{fig:etas}, with $\xi_k/S_k$ and $\xi_r/\sigma^2$
approaching 1 at small $d$ values, and the points becoming
uncorrelated at large distances. Note also that $\xi_k/S_k$ again
contains small oscillations at intermediate distances.

\subsection{Two Correlated Random Walks}

With these correlation functions in hand, we now consider the
simultaneous correlated random walks of two overdensities
$\del_1(S_{k,1})$ and $\del_2(S_{k,2})$ separated by a fixed
Lagrangian distance $d$. As in the one-point case, for the derivation
we adopt sharp $k$-space filters. We want to determine the joint probability
distribution of these two densities,
$Q(\del_1,\del_2,S_{k,1},S_{k,2},d).$
In terms of a trajectory density in the $(\del_1,\del_2)$ plane,
$Q(\del_1,\del_2,S_{k,1},S_{k,2},d)\, d\del_1\, d\del_2$ is the
fraction of trajectories that are in the interval $\del_1$ to
$\del_1+d\del_1$ and $\del_2$ to $\del_2+d\del_2$ at $(S_{k,1},
S_{k,2})$.  Below we will take $S_{k,1}$ and $S_{k,2}$ to be the {\em
final}\, variances of these trajectories, denoting intermediate
variances with the primed notation $S'_{k,1}$ and $S'_{k,2}.$ We then
consider a large number of random walks all of which begin with
$\del_1=0$ and $\del_2=0$ at $S'_{k,1}=0$ and $S'_{k,2}=0$.

With sharp $k$-space filters, the problem simplifies due to the fact
that we are working with a Gaussian random field. Suppose we consider
random walks $\del_1(S_{k,1}')$ and $\del_2(S_{k,2}')$ over the ranges
$0 \leq S_{k,1}' \leq S_{k,1}$ and $0 \leq S_{k,2}' \leq
S_{k,2}$. Since different $k$-modes are independent of each other, any
part of the random walk $\del_1$, e.g., in the range $S_k < S_{k,1}' <
S_k+dS_k$, is only correlated with the same range, $S_k < S_{k,2}' <
S_k+dS_k$, in the other random walk, with a correlation strength
determined by $\eta(d,S_k)$. This means that in order to determine the
joint probability distribution $Q(\del_1,\del_2,S_{k,1},S_{k,2},d)$,
we do not need to vary $S_{k,1}'$ and $S_{k,2}'$ independently;
instead we can consider $Q$ to be a function of a single variable
$S'_k$ (in addition to the variables $\del_1$, $\del_2$, and $d$). If
the final $S'_k$ values are unequal, e.g., $S_{k,1} > S_{k,2}$, then
we continuously increase $S'_k$ from 0 to $S_{k,2}$, generating the
two correlated random walks and solving for $Q$ on the line $S'_{k,1}
= S'_{k,2} = S'_k$ in the $(S_{k,1},S_{k,2})$ plane, and we then
continue to increase $S'_k$ up to $S_{k,1}$, with only the random walk
$\del_1$ continuing further (with $S_{k,1}'=S'_k$).

We thus divide the evolution of $Q$ into two segments, one in which
the $k$-space filter includes only small wavenumbers, such that $S'_k$
is smaller than both $S_{k,1}$ and $S_{k,2}$, and a second segment of
wavenumbers such that $S'_k$ is between $S_{k,1}$ and $S_{k,2}$. With
$d$ fixed, we follow the derivation from the one-point case and
expand $Q(\del_1 - \Delta \del_1, \del_2 - \Delta \del_2,S'_k -
\Delta S'_k,d)$ to obtain
\be \Delta S'_k \frac{\partial Q} {\partial S'_k} =
\frac{1} {2} \left\langle\Del \del_1^2\right\rangle \frac{\partial^2
Q} {\partial \del_1^2} +  \left\langle\Del \del_1 \Del
\del_2 \right\rangle \frac{\partial^2 Q}{\partial \del_1 \del_2}+
\frac{1} {2} \left\langle\Del \del_2^2\right\rangle \frac{\partial^2
Q} {\partial \del_2^2}. \ee Evaluating the expectation values in each
regime gives
\be
\frac{\partial Q} {\partial S'_k} =
\cases {
\frac{1} {2}  \frac{\partial^2 Q} {\partial \del_1^2} +
\eta(d,S'_k)\,
\frac{\partial^2 Q} {\partial \del_1 \del_2}+
\frac{1}{2} \frac{\partial^2 Q} {\partial \del_2^2}
	& $\ S'_k < S_{k,1}, S_{k,2}$ \cr
\frac{1}{2}  \frac{\partial^2 Q} {\partial \del_1^2}
	& $\ S_{k,2} < S'_k < S_{k,1}$ \cr
\frac{1}{2}  \frac{\partial^2 Q} {\partial \del_2^2}
	& $\ S_{k,1} < S'_k < S_{k,2}$\ .
\cr} \ee

In order to compute the evolution of these points in the regime in
which $S'_k < S_{k,1}, S_{k,2}$ it is simpler to transform to the
uncorrelated variables $\del_+ = (\del_1 + \del_2)/\sqrt{2}$ and
$\del_- = (\del_1 - \del_2)/\sqrt{2}$.  In this case we obtain
\be
\frac{\partial Q} {\partial S'_k} =
\frac{1 + \eta}{2}\,
        \frac{\partial^2 Q} {\partial \del_+^2} +
\frac{1 - \eta}{2}\,
        \frac{\partial^2 Q} {\partial \del_-^2} \ .
\label{eq:two}
\ee
In these variables the two random walks are independent, and the usual
no-barrier solution is
\be
Q_0(\del_+, \del_-, S_{k,{\rm min}},\xi_k) =
G[\del_+,S_{k,{\rm min}} + \xi_k]\, G[\del_-,S_{k,{\rm min}} - \xi_k]\ ,
\label{eq:diff2}
\ee
where
\be
S_{k,{\rm min}} \equiv {\rm min}(S_{k,1},S_{k,2})\ ,
\ee
and the
covariance of $\del_1$ and $\del_2$ at $S_{k,{\rm min}}$ is $\xi_k \equiv
\xi_k(d,S_{k,{\rm min}})$ [see eq.~(\ref{eq:xi})]. The solution at the 
point $(S_{k,1},S_{k,2})$ is then simply obtained by convolving
$Q_0(\del_+, \del_-, S_{k,{\rm min}},\xi_k)$ with $G(\del_1,S_{k,1} -
S_{k,2})$ or $G(\del_2,S_{k,2} - S_{k,1}).$

>From eq.~(\ref{eq:diff2}) we see that in the $\del_+$ and $\del_-$
coordinates, the evolution of the densities is particularly simple,
and there are clearly image solutions of the form $Q_0(w_+ -
\del_+,w_- - \del_-,S_{k,{\rm min}},\xi_k)$, where $w_+$ and $w_-$ are
arbitrary constants. But unlike the one-dimensional case, the
absorbing barriers at $\del_1=\nu_1$ and $\del_2=\nu_2$ are
complicated due to the coordinate transformation, where, e.g., one of
the barriers takes the form $\del_+\, +\, \del_- = w$, where $w$ is a
constant. We want a solution for $Q$ which is zero on this barrier,
but the image solutions do not help since $\del_+$ and $\del_-$ are
coupled in the equations that describe the barriers. An exact solution
thus requires a numerical approach, in which the diffusion equation is
solved with the absorbing barrier imposed at each time step. However,
in the following section we derive an accurate analytic approximation
to the exact solution.

\subsection{Two-Step Approximation}

While the full solution of the double barrier problem requires a
numerical approach, a closer look at the problem leads to a a simple
approximate analytic solution that captures the underlying physics of
two-point collapse. Note that we need only approximate the evolution
for $0 \leq S'_k \leq S_{k,{\rm min}}$, since the evolution for higher
values of $S'_k$ involves only one of the random walks and therefore
has a simple exact solution.

We consider the equation for the differential correlation
function between the halos as a function of mass, eq.\ (\ref{eq:etak}).
While $\eta(d,S'_k)$ is an oscillating function, it equals unity at small
values of $S_k$ and its amplitude steadily declines with $S'_k$ as the
corresponding wavenumber $k$ enters the regime in which $k r \gg 1$. Thus
for small $S'_k$ values, the two random walks are essentially identical,
with $\del_1(S'_k) \simeq \del_2(S'_k)$. In this regime we can simply evolve a
single random walk in $\del_1$, and then set $\del_2$ equal to the
resulting final value of $\del_1$. At large $S'_k$, on the other hand,
$\eta(d,S'_k) \rightarrow 0$, the two random walks become independent,
and the problem again simplifies.

\begin{figure}
\plotone{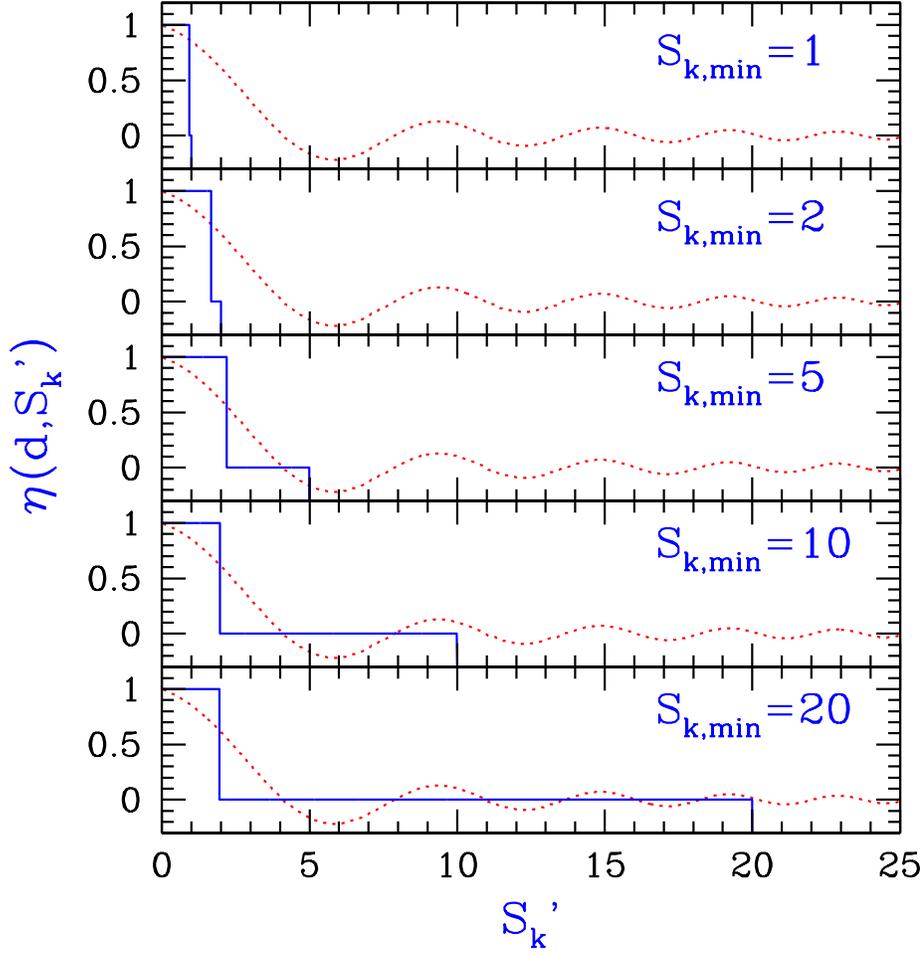}
\caption{Comparison 
between $\eta(d,S'_k)$ calculated exactly (solid lines) and in the
two-step approximation (dotted lines), for various values of
$S_{k,{\rm min}}$.  In all plots $d = 3.3$ comoving $\hMpc$.}
\label{fig:approx}
\end{figure}

These observations lead us to propose a ``two-step'' approximation in
which we replace $\eta(d,S'_k)$ with a simple step function that jumps
from unity to zero at some value of $S'_k$.  We choose this jump to
occur at a value of $S'_k$ that preserves the exact solution for $Q$
at $S'_k=S_{k,{\rm min}}$ in the absence of the barriers. In the
no-barrier case, the joint distribution of $\del_1$ and $\del_2$ at
$S_{k,{\rm min}}$ depends only on the variance $S_{k,{\rm min}}$ of
each of these Gaussian variables, and on their covariance
$\xi_k(d,S_{k,{\rm min}})$ which is given by eq.\ (\ref{eq:xi}). In
order for the integral in eq.\ (\ref{eq:xi}) to be the same for both
the exact function $\eta(d,S'_k)$ and for the step-function
approximation, we must fix the step to occur at $S'_k
=\xi_k(d,S_{k,{\rm min}})$. Thus, our two-step approximation is
\be 
\eta(d,S'_k) \simeq \cases { 1 & $0 \leq
S'_k \leq \xi_k(d,S_{k,{\rm min}})$ \cr 
0 & $\xi_k(d,S_{k,{\rm min}}) < S'_k
\leq S_{k,\rm min} $\ , \cr} 
\ee 
such that the trajectories are completely correlated when $S'_k$ is less
than the cross-correlation between the points at $S_{k,{\rm min}}$,
and completely uncorrelated when $S'_k$ exceeds this value. This
approximation is compared to the exact form of $\eta(d,S'_k)$ for
various values of $S'_k$ in Fig.\ \ref{fig:approx}.

We combine this approximation for the evolution of the two random
walks for $S'_k \leq S_{k,{\rm min}}$ with the single random walk which
continues for $S'_k > S_{k,{\rm min}}$, resulting in the following overall
prescription with absorbing barriers at $\del_1=\nu_1$ and
$\del_2=\nu_2$. First, we evolve $\del_1$ for $0 \leq S'_k \leq
\xi_k(d,S_{k,{\rm min}})$. Since we are assuming that the two random walks
are identical in this regime, we must place the barrier on $\del_1$ at
\be
\nu_{\rm min} \equiv {\rm min}(\nu_1,\nu_2)\ .
\ee
Quantitatively, the single absorbing barrier solution,
eq.\ (\ref{eq:1}), gives $Q$ at $S'_k=\xi_k(d,S_{k,{\rm min}})$,
\be
Q_a(\nu_{\rm min},\del_1,\del_2, \xi_k) =
[G(\del_1, \xi_k)-G(2 \nu_{\rm min} -  \del_1, \xi_k)] \,
\del_D(\del_1 - \del_2) \theta(\nu_{\rm min} - \delta_1),
\label{eq:Qa}
\ee
where $\del_D$ is a one-dimensional Dirac delta function, $\theta$ is
the Heaviside step function, and here and in the rest of this section
$\xi_k$ refers to $\xi_k(d,S_{k,{\rm min}})$.  We then set $\del_2 =
\del_1$ and evolve the random walks in $\del_1$ and $\del_2$
independently up to $S_{k,1}$ and $S_{k,2}$, with the barriers at
$\nu_1$ and $\nu_2$, respectively. Thus, we first convolve eq.\
(\ref{eq:Qa}) with the no-barrier solutions for the two independent
random walks,
\be
Q_{b}(\del_1,\del_2,S_{k,1},S_{k,2}, \xi_k) =
G(\del_1,S_{k,1}- \xi_k)\, G(\del_2,S_{k,2} - \xi_k)\ ,
\ee
which gives
\ba
Q_0(\nu_{\rm min},\del_1,\del_2,S_{k,1},S_{k,2},\xi_k)
&=& Q_+(\nu_{\rm min},\del_1,\del_2,S_{k,1},S_{k,2},\xi_k)\ + \nonumber \\
&& Q_-(\nu_{\rm min},2 \nu_{\rm min} - \del_1,2 \nu_{\rm min} -
\del_2,S_{k,1},S_{k,2},\xi_k)\ ,
\label{eq:q0two}
\ea
where
\ba
Q_{\pm} (\nu_{\rm min},
\del_1,\del_2,S_{k,1},S_{k,2},\xi_k) & \equiv&  \frac{1}{4 \pi \sqrt{S_{k,1} 
S_{k,2} -
\xi_k^2}} \, \exp \left[ -\, \frac{\del_1^2 S_{k,2}
+  \del_2^2 S_{k,1} -
2 \del_1 \del_2 \xi_k} {2 (S_{k,1} S_{k,2} - \xi_k^2)}
\right]  \nonumber \\
& & \times \left[{\rm erf} \left( \tilde \nu
\sqrt{ \frac{\tilde S}{2}}\right) \pm 1 \right],
\label{eq:Qpm}
\ea
and
\be
\tilde S \equiv
\frac{ \xi_k (S_{k,1} - \xi_k) (S_{k,2} - \xi_k)}
{S_{k,1} S_{k,2} - \xi_k^2}\ , \qquad \qquad
	\tilde \nu \equiv \
	\frac{\nu_{\rm min}}{\tilde S} - \frac{\del_1}{S_{k,1} - \xi_k}
	 -  \frac{\del_2}{S_{k,2} - \xi_k}\ .
\ee

Finally, we account for the additional barriers with reflections
about the $\nu_1$ and $\nu_2$ axes, which yields
\ba
Q(\nu_1,\nu_2,\del_1,\del_2,S_{k,1},S_{k,2},\xi_k)
& = & Q_0(\nu_{\rm min},\del_1,\del_2,S_{k,1},S_{k,2},\xi_k)\ +
	\nonumber \\
&& Q_0(\nu_{\rm min},2\nu_1-\del_1,2\nu_2-\del_2,S_{k,1},S_{k,2},\xi_k)\ -
	\nonumber \\
&& Q_0(\nu_{\rm min},\del_1,2\nu_2-\del_2,S_{k,1},S_{k,2},\xi_k)\ -
	 \nonumber \\
&& Q_0(\nu_{\rm min},2\nu_1-\del_1,\del_2,S_{k,1},S_{k,2},\xi_k)\ .
\label{eq:approxanswer}
\ea
Following the common approximation taken in the single particle case
we again replace $S_{k,1}$ and $S_{k,2}$ with $\sigma^2(M_1)$ and
$\sigma^2(M_2)$, respectively. Similarly, we also replace $\xi_k$ with
the correlation function of real-space peaks $\xi_{r_{max}}$ as given
by eq.\ (\ref{eq:xirmax}).  Using this expression we can now construct
the combined mass function of halos at two points separated by a
comoving Lagrangian distance $d$. Before we derive this function,
however, it is important to understand the errors introduced by our
simple two-step approach. Thus, we compare eq.\
(\ref{eq:approxanswer}) with exact numerical solutions in order to
show that our analytic approximation is accurate over a broad range of
parameter space.

\subsection{Numerical Approach and Comparison with Two-Step
Approximation}

In order to solve eq.\ (\ref{eq:two}) in the presence of absorbing
barriers at fixed values of $\del_1$ and $\del_2$ we have developed a
simple finite difference code.  We construct a 400 $\times$ 400 zone
mesh in $\del_+$ and $\del_-$ spanning the range from $- 5\,
\delta_{\rm scale} \leq \del_+ \leq 5\, \delta_{\rm scale}$ and $- 5\,
\delta_{\rm scale} \leq \del_- \leq 5\, \delta_{\rm scale}$, where
$\delta_{\rm scale}$ is a typical overdensity of interest.  In this
case the width of each zone, $dx$, is $0.025\, \delta_{\rm scale}$ in
both dimensions. On this grid, we construct $Q_{i,j}^{t} =
Q(\del_{+,i},
\del_{-,j},S'_k(t))$ where $i$ and $j$ are spatial indices in each of
directions and $t$ is a ``time'' counter such that $S'_k(t) = t\,
dS'_k,$ where we take $dS'_k = \delta_{\rm scale}^2/3000$ to be the
interval by which we refine our $k$-space filter at each time step.

Initially the distribution is taken to be a delta function, such that
$Q^0_{0,0} = 1$ and $Q^0_{i,j} = 0$ for all other values of $i$ and
$j$.  We calculate the values at each new step in $S'_k$ using an
alternating-direction implicit method \citep{p92} in which we divide
each time step into two stages of size $dS'_k/2$, and solve first in
the $x$ and then the $y$ direction.  In this case eq.\ (\ref{eq:two})
becomes
\ba
Q^{t+1/2}_{i,j} &=& Q^{t}_{i,j}  +
\frac{\alpha_+}{2}
\left[Q^{t+1/2}_{i-1,j}-2 Q^{t+1/2}_{i,j}+Q^{t+1/2}_{i+1,j}\right]  +
\frac{\alpha_-}{2}
\left[Q^{t}_{i,j-1}   -2 Q^{t}_{i,j} + Q^{t}_{i,j+1}\right], \nonumber \\
Q^{t+1}_{i,j} &=& Q^{t+1/2}_{i,j}  +
\frac{\alpha_+}{2}
\left[Q^{t+1/2}_{i-1,j}-2 Q^{t+1/2}_{i,j}+Q^{t+1/2}_{i+1,j}\right]  +
\frac{\alpha_-}{2}
\left[Q^{t+1}_{i,j-1} -2 Q^{t+1}_{i,j} + Q^{t+1}_{i,j+1}\right],
\ea
where $\alpha_+ \equiv \frac{[1 +
\eta(d,S'_k)]dS'_k}{2\, d x^2}$, $\alpha_- \equiv \frac{[1 -
\eta(d,S'_k)]dS'_k}{2\, d x^2}$, and the system can be solved
using simple inversions of tridiagonal matrices.
Finally, at the end of each time
step we impose the absorbing barriers at $\del_1 = \nu_1$ and
$\del_2 = \nu_2 $ by setting $Q_{i,j}^{t+1,n} = 0$ at all points at
which $\del_{+,i} +\del_{-,j} \geq \sqrt{2}\, \nu_1$ or $\del_{+,i}
-\del_{-,j} \geq \sqrt{2}\, \nu_2.$
With this code we are able to examine the accuracy of our two-step
approximation for a number of physical cases.

\begin{figure}
\plotone{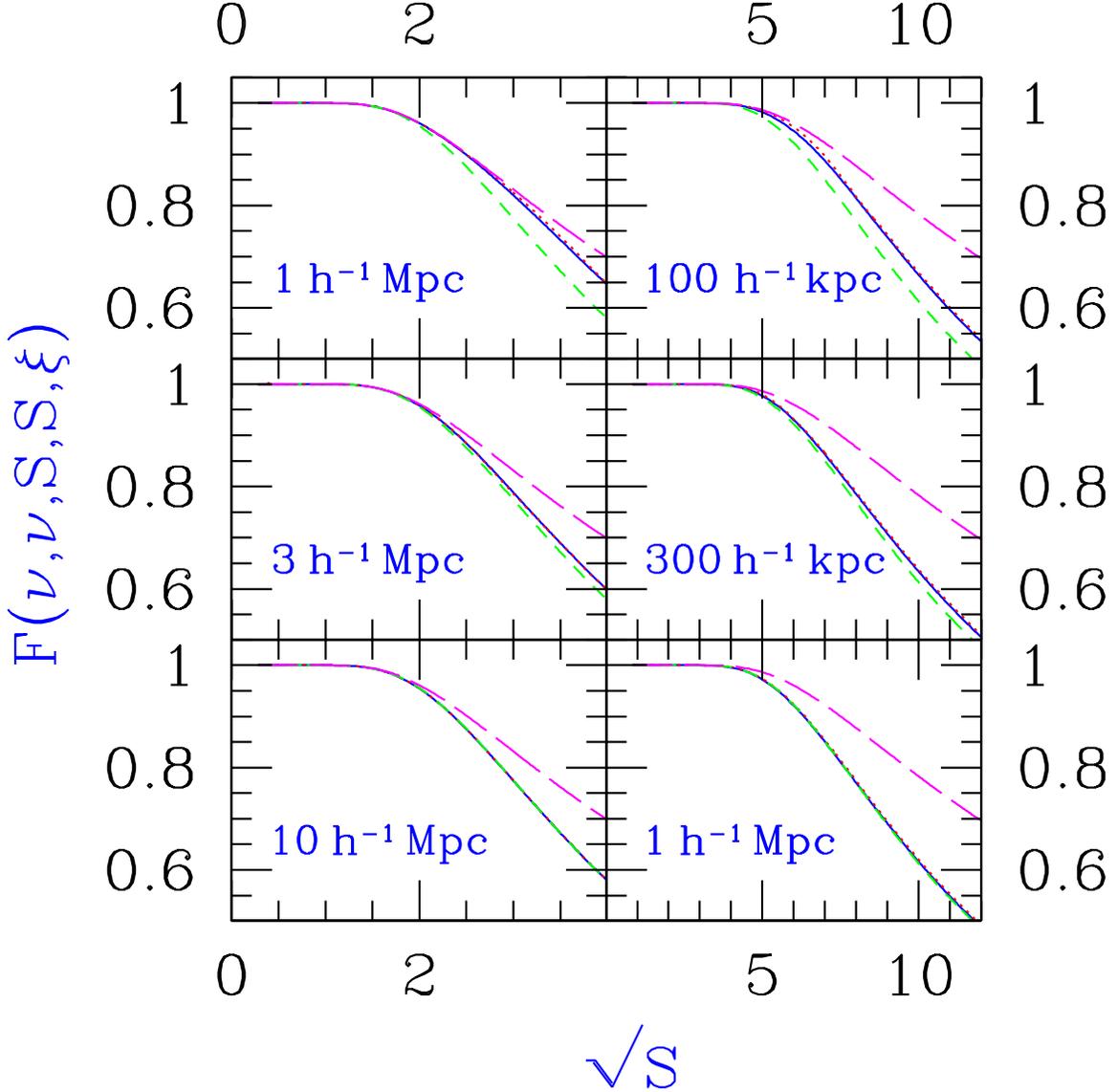}
\caption{Comparison between numerical solution and two-step
approximation as a function of $S$.  In the left panels we consider
two peaks that collapse at redshifts 2 and 3, respectively, while in
the right panels we consider two peaks that collapse at $z = 8$. The
panels are labeled by the comoving Lagrangian distance between the
peaks, and in each panel the numerical solution and the two-step
solution are given by the solid and dotted line, respectively, while
for comparison, the fully uncorrelated solution is given by the
short-dashed line and the fully correlated solution is given by the
long-dashed line. Note that the two-step approximation is almost
indistinguishable from the numerical solution, in every panel. Note
also that in this plot we use $S_k$ for $S$ and $\xi_k$ for $\xi$.}
\label{fig:Scompare}
\end{figure}

\begin{figure}
\plotone{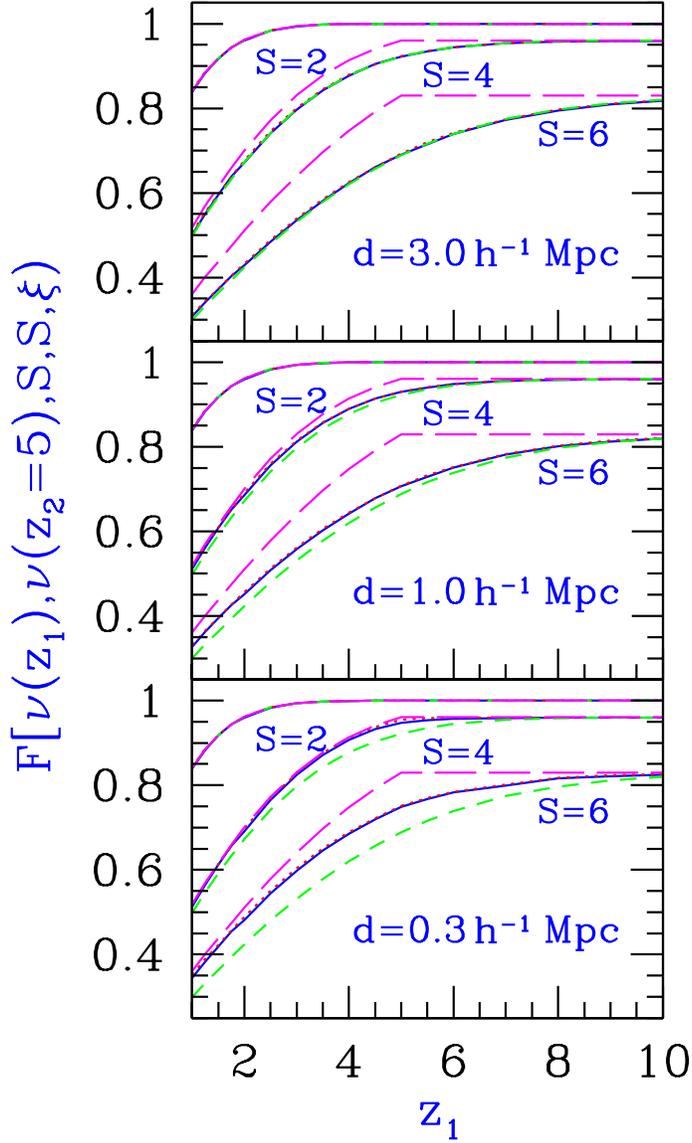}
\caption{Comparison between numerical solution and two-step
approximation as a function of $z_2$.  The panels are again labeled by
the comoving Lagrangian distance between the peaks, and in each panel
the numerical solution and two-step solutions are given by the solid
and dotted line, respectively, and the fully uncorrelated and fully
correlated solutions are given by the short-dashed and long-dashed
line, respectively. All four lines coincide for $S = 2$. For $S = 4$
and $S = 6$, our approximation and the numerical calculation are
almost indistinguishable, but the uncorrelated and the fully
correlated lines lie significantly lower and higher,
respectively. Note that in this plot we use $S_k$ for $S$ and $\xi_k$
for $\xi$.}
\label{fig:nucompare}
\end{figure}

For any given redshifts $z_1$ and $z_2$, we fix the barriers  at the
collapse thresholds $\nu_1=\del_c(z_1)$ and $\nu_2=\del_c(z_2)$, and
consider the following quantity:
\be
F(\nu_1,\nu_2,S_{k,1},S_{k,2},\xi_k)=\int_{-\infty}^{\nu_1} d\del_1
\int_{-\infty}^{\nu_2} d\del_2\,
Q(\nu_1,\nu_2,\del_1,\del_2,S_{k,1}, S_{k,2},\xi_k)\ .
\label{eq:bigF}
\ee 
This is the fraction of trajectories that are not absorbed before the
random walks reach the point $(S_{k,1}, S_{k,2})$.  Recall that $S_k$
can be related to the mass scale of the collapsed objects by the
simple ansatz $S_k \rightarrow \sigma^2(r)$ as described in \S 3.1
Thus equation eq.\ (\ref{eq:bigF}) can be interpreted as the
probability of point $A$ being in a halo of mass $M <
M_1(\sigma^2=S_{k,1})$ and point $B$ in a halo of mass $M <
M_2(\sigma^2=S_{k,2})$. Since $F$ is the basic quantity that we later
use to calculate various halo properties, a reasonable way to judge
the success of our approximate method is to test its ability to
closely reproduce this function. In the comparisons
(Figures~\ref{fig:Scompare} and \ref{fig:nucompare}), we use
the $k$-filter quantities $S_k$ and $\xi_k$. 

In Figure~\ref{fig:Scompare} we compare analytical and numerical
values of $F(\nu_1,\nu_2,S_{k,1},S_{k,2},\xi_k)$ as a function of
$S_{k,1} = S_{k,2} = S$ for various distances and collapse redshifts.
In the left three panels of this figure, we consider large objects
collapsing at different redshifts by fixing $\nu_1 = 4.13 =
\delta_c(z_1=2)$, $\nu_1 = 5.47 = \delta_c(z_2=3)$, $\delta_{\rm
scale} =3.0$ and varying the distances as labeled.  In the right
panels we consider the simultaneous collapse of somewhat smaller
objects at earlier times fixing $\nu_1 = \nu_2 = 12.35 = \delta_c
(z=8)$ and $\delta_{\rm scale} = 8.0$. The figure shows that both for
peaks that form simultaneously, and for peaks that collapse at different
times, the two-step solution tracks the numerical solution over a large
range of halo masses and distances. Note also that these expressions
match each other closely even when they are very different from both
the correlated and the uncorrelated cases. 

In Figure \ref{fig:nucompare} we fix three values of $S_{k,1} =
S_{k,2} = S$ and the collapse redshift of a single halo $z_1 = 5$, and
consider $F$ as a function of the collapse redshift $z_2$ of the
second halo, at three different separations. The figure shows that our
approximation does well at reproducing the numerical results even for
cases in which the collapse redshifts between the two objects are very
different. As before, the numerical and approximate results are in
good agreement over a wide range of values for which the
fully-correlated and fully-uncorrelated expressions are not accurate.

Besides the results shown in these figures, we have conducted
extensive convergence tests varying $dS$ and $dx$, and have found that
reducing any of these parameters has no effect on our numerical
results. Thus we are confident that the differences between the
analytic and exact solutions, as presented in Figures
\ref{fig:Scompare} and \ref{fig:nucompare}, are in general small, and the
errors in our analytical expressions are at most $2 \%.$

\section{Structure Formation with the Two-Point Formalism}

\subsection{Bivariate Mass Function}

Having developed in the previous section an accurate approximation to
the joint probability distribution at two points, we now apply this
distribution to derive the two-point generalization of the
Press-Schechter mass function of collapsed halos.  This generalization
is the joint probability of having point $A$ lie in a halo in the mass
range $M_1$ to $M_1 + dM_1$ at a redshift $z_1$ and point $B$ lie in a
halo in the mass range $M_2$ to $M_2 + dM_2$ at a redshift $z_2$.
Note that the derivation in $\S 3.2$ applies to any redshifts $z_1$
and $z_2$ and thus our calculation can determine the number density of
halos at $B$ both before and after the formation of a halo at $A$.

Again, it is important to emphasize that this joint mass function is
defined in the Lagrangian coordinate system that arises naturally in
an excursion-set approach.  Thus the distance between the points in
physical space may be somewhat different than the $d$ considered in
our equations, as the relevant comoving distance in our case is that
separating the points at early times.  This coordinate system has both
its weaknesses and its advantages.  While it somewhat complicates the
direct comparison of our expressions with numerical simulations, this
can in general be remedied with estimates of the final, Eulerian halo
coordinates \citep[e.g.,][]{mo96}.  Furthermore, for many
applications, Lagrangian results are in fact preferable to an Eulerian
description in which halos are indexed by their physical coordinates
with no reference to where these perturbations came from. In studies of
spatially-dependent feedback in structure formation, for example, it
is often more important to have a measure of the total column depth of
material separating two perturbations than their precise distance in
physical space.

In this section, we adopt a general notation for the variances and
correlation functions, using $S$ to represent either the $k$-space
filtered quantity, $S_k$, its real space equivalent $\sigma^2(r)$, or
any alternative definition.  Similarly, $\xi$ denotes $\xi_{r_{max}}$,
$\xi_{k(r)}$, or any alternative definition of the correlation
function.  With this notation, we now consider
$f(\nu_1,\nu_2,S_1,S_2,\xi)\, dS_1\, dS_2$, the probability of having
point $A$ in a halo with mass corresponding to the range $S_1$ to
$S_1+d S_1$ and point $B$ in a halo with mass corresponding to the
range $S_2$ to $S_2+d S_2$. This is simply related to the quantity in
eq.\ (\ref{eq:bigF}) as
\be
f(\nu_1,\nu_2,S_1,S_2,\xi) = \frac{\partial}{\partial S_1}
\frac{\partial}{\partial S_2} \int_{-\infty}^{\nu_1} d\del_1
\int_{-\infty}^{\nu_2} d\del_2\, Q(\nu_1,\nu_2,\del_1,
\del_2,S_1,S_2,\xi)\ , 
\ee 
where $\xi$ is not considered an
independent variable (and so the partial derivatives involve
variations of $\xi$). This simplifies to 
\be 
f(S_1,S_2,\nu_1,
\nu_2,\xi) = \frac{\partial}{\partial S_1} \frac{\partial} {\partial
S_2} \left[2 \int_{-\infty}^{\nu_1} d\del_1 -
\int_{-\infty}^{\infty} d \del_1 \right] \left[2
\int_{-\infty}^{\nu_2} d\del_2 - \int_{-\infty}^{\infty} d \del_2
\right] Q_0(\nu_1,\nu_2,\del_1,\del_2,S_1,S_2,\xi)\ .
\label{eq:f1} 
\ee

In order to perform these integrals, we must consider $Q_0$ from eq.\
(\ref{eq:q0two}) in its unintegrated form, written as a convolution
with an integration variable $x$: \be Q_0 = \int_{-\infty}^{\nu_{\rm
min}} dx\, \left[ G(x, \xi) - G(2 \nu_{\rm min} - x, \xi)
\right]\, G(\del_1 - x,S_1- \xi)\, G(\del_2 - x, S_2 - \xi)\
. \label{eq:dxQ0} \ee
Since $S_1$ and $\del_1$ appear only in a single term in the
$x$-integration, and similarly for $S_2$ and $\del_2$, $Q_0$ satisfies
\be
\frac{\partial Q_0}{\partial S_1} = \frac{1}{2}\frac{\partial^2
Q_0}{\partial \del_1^2} + \frac{\partial \xi} {\partial S_1}
\frac{\partial Q_0}{\partial \xi}\ , \qquad {\rm and} \qquad
\frac{\partial Q_0}{\partial S_2} = \frac{1}{2}\frac{\partial^2
Q_0}{\partial \del_2^2} + \frac{\partial \xi} {\partial S_2}
\frac{\partial Q_0}{\partial \xi}\ ,
\ee
where we consider $\xi$ to be an independent variable
when calculating $\frac{\partial Q_0}{\partial \xi}.$

We perform the partial derivatives with respect to $S_1$ and $S_2$ on
the integrand in equation (\ref{eq:dxQ0}), obtaining four terms. Terms
that contain partial derivatives with respect to $\del_1$ or $\del_2$
allow us to perform the $\del_1$ or $\del_2$ integrations in eq.\
(\ref{eq:f1}) trivially, while the evaluation of terms containing
$\frac{\partial Q_0}{\partial \xi}$ is more involved.

To compute $\frac{\partial Q_0}{\partial \xi}$, we note that the
integrand in eq.\ (\ref{eq:dxQ0}) is a product of three terms, which
we denote, respectively from left to right, as $I_1$, $I_2$, and
$I_3$. These satisfy the equations
\be
\frac{\partial I_1}{\partial \xi} =
\frac{1}{2}\frac{\partial^2 I_1}{\partial x^2}\ , \qquad {\rm and}
\qquad \frac{\partial I_{2,3}}{\partial \xi} =
-\frac{1}{2}\frac{\partial^2 I_{2,3}}{\partial x^2},
\ee
which allow us to perform the integration with respect to $x$, using
integration by parts to eliminate all double derivatives.  
This yields an integrated term, whose
contribution to $f(\nu_1,\nu_2,S_1,S_2,\xi)$ is zero, and the
remaining term in 
$\frac{\partial Q_0}{\partial \xi}$
is $$
\int_{-\infty}^{\nu_{\rm min}} dx \, I_1\, \partial_x I_2\, \partial_x
I_3\ . $$ Finally we use $\partial_x I_2 = - \partial_{\del_1} I_2$ and
$\partial_x I_3 = -\partial_{\del_2} I_3$ to further simplify any term
that contains a partial derivative with respect to $\xi$, yielding
\be
f(\nu_1,\nu_2,S_1,S_2,\xi) = \left. \left[ \frac{\partial^2 Q_0}
{\partial \del_1 \partial \del_2} +\, 2 \left( \frac{\partial \xi}
{\partial S_1}\, \frac{\partial^2 Q_0} {\partial \del_2^2} +
\frac{\partial \xi} {\partial S_2}\, \frac{\partial^2 Q_0} {\partial
\del_1^2} \right) + 4 \left( \frac{\partial \xi} {\partial S_1}
\frac{\partial \xi} {\partial S_2} \frac{\partial Q_0} {\partial \xi}
+\frac{\partial^2 \xi}{\partial S_1 \partial S_2}\, Q_0 \right)
\right] \right|_{\del_1 = \nu_1,\, \del_2 = \nu_2}\ .
\label{eq:littlef}
\ee

Note that this expression is discontinuous for some choices of $S$ and
$\xi$. This is true both for the $k$-space-filtering case for which
eqs.\ (\ref{eq:Qpm}) and (\ref{eq:approxanswer}) were derived, as well
as the real-space case in which $\xi = \xi_{r_{max}}$ and $S =
\sigma^2$, because in each case $\xi$ is a function of $S_{\rm min}$
whose derivatives with respect to $S_1$ and $S_2$ are not continuous.
While this may seem at first to be a serious problem, a closer look at
these discontinuities shows that they are in fact quite harmless.  

To see why this is true, consider the case in which $S_1$ and $S_2$
are nearly equal. If $S_1=S_2-\epsilon$, then only the
$\frac{\partial^2 Q_0}{\partial \delta_1 \partial \delta_2}$ and
$\frac{\partial^2 Q_0} {\partial \delta_2^2}$ terms of eq.\
(\ref{eq:littlef}) are nonzero (since in this case $\xi$ is a function
of $S_{\rm min}=S_1$), while if $S_1=S_2+\epsilon$ then the only
nonzero terms are the mixed derivative term and $\frac{\partial^2
Q_0}{\partial \delta_1^2}$. Now, in the limited case in which
$\nu_1=\nu_2$, $\frac{\partial^2 Q_0}{\partial \delta_1^2}$ and
$\frac{\partial^2 Q_0}{\partial \delta_2^2}$ are equal at $S_1=S_2$
and $f$ is continuous at this point, while if $\nu_1 \neq \nu_2$, $f$
is discontinuous at $S_1=S_2$.

This discontinuity at $\nu_1 \neq \nu_2$ is not a limitation of our
method, however, but is instead a true reflection of the physical
situation. This is clearest in the $d = 0$ case in which the two
points coincide. In this case, if $\nu_1 < \nu_2$ and thus $z_1 <
z_2$, taking $S_1 = S_2 - \epsilon$ corresponds to $M_1 > M_2$ which
is simply the accretion of mass over time. Taking $S_1 = S_2 +
\epsilon,$ however, would correspond to {\em losing}\, mass with time,
which is contradictory to our most basic assumptions. Thus,
approaching $S_1 = S_2$ from different directions has two different
meanings, only one of which is relevant to structure formation. The
problem therefore arises from defining an effective arrow of time and
considering only the first-crossing distribution with respect to
it. Thus, the discontinuity only occurs in transitions between
physically relevant choices of parameters and regions of parameter
space that contradict the whole premise of our approach, and it need
not concern us.

In Appendix A we give all the expressions necessary
to evaluate eq.\ (\ref{eq:littlef}) explicitly. We can thus
analytically compute the joint halo abundance as
\be
\frac{d^2 n^2_{12}}{dM_1 dM_2} =
\frac{\bar{\rho}}{M_1} \left|\frac{d S_1}{d M_1} \right|
\frac{\bar{\rho}}{M_2} \left|\frac{d S_2}{d M_2} \right|
f(\nu_1,\nu_2,S_1,S_2,\xi)\ ,
\label{eq:2ptm}
\ee which reduces to the product of $\frac{dn_1}{dM_1}$ and
$\frac{dn_2}{dM_2}$ as given by eq.\ (\ref{eq:abundance}) in the limit
of large distances, when $\xi \rightarrow 0$.  Note that the last two
terms in eq.~(\ref{eq:littlef}) are identically zero if we use
$\xi_k$ or $\xi_{r_{max}}$ as our definition of $\xi$.

\begin{figure}
\plotone{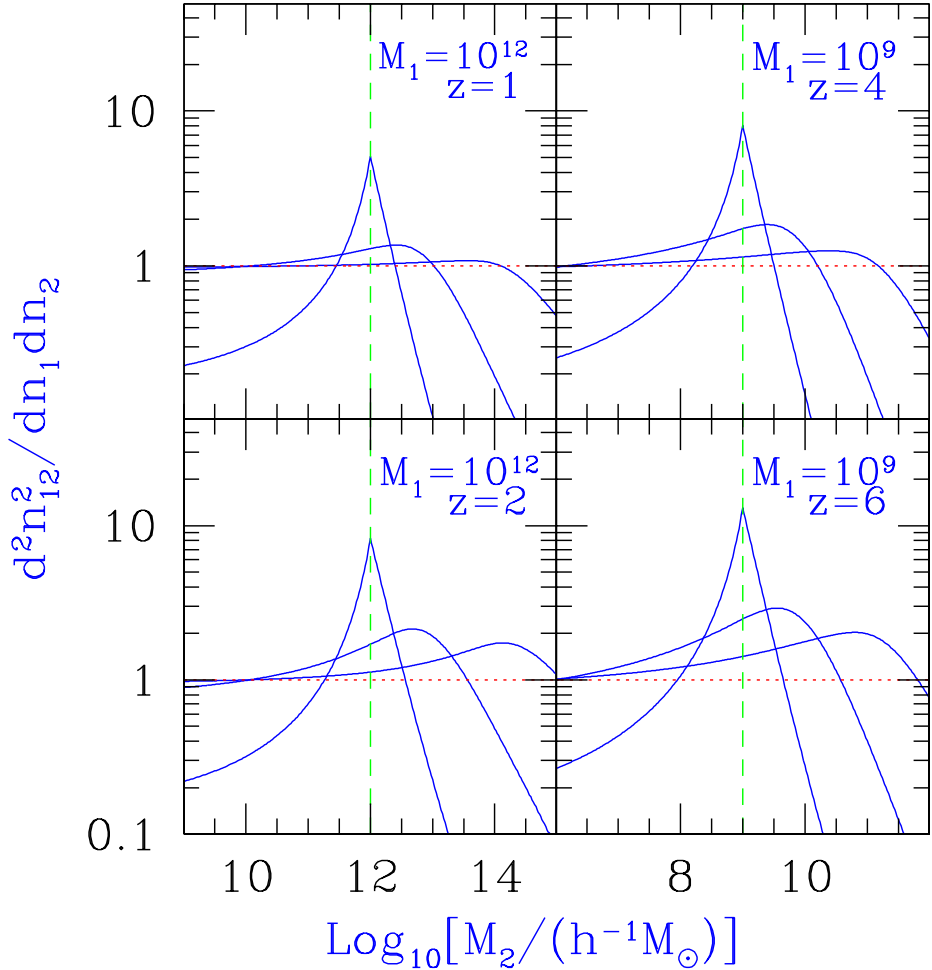}
\caption{Normalized number density of correlated halos as calculated
from eq.\ (\ref{eq:2ptm}) as a function of $M_2$, for various values
of distance, $M_1$, and redshift. In the left panels $M_1$ is held
fixed at a value of $10^{12} \hMsun$, roughly corresponding to an
$L_\star$ galaxy, at redshifts of 1 and 2.  Here the solid lines show,
from top to bottom (in terms of the highest point in each curve),
results with the Lagrangian distances between the two points fixed at
1.0, 3.3, and 10 comoving $\hMpc$ respectively, while for reference
the uncorrelated case is plotted as the dotted line at 1
and the mass at which $M_1 = M_2$ is given by the vertical 
dashed line.  In the right
panels, we fix $M_1$ at $10^9 \hMsun$ corresponding to a dwarf galaxy.
Here we consider a higher range of redshift values ($z_1= z_2 = 4$ and
$z_1= z_2 = 6$) roughly corresponding to the same values of
$\nu/\sigma$ used in the $M_1 = 10^{12} \hMsun$ case.  In these panels
the solid lines show normalized densities at $d$ = 0.1, 0.33, and 1.0
comoving $\hMpc$, from top to bottom.}
\label{fig:ps2}
\end{figure}

In Figure \ref{fig:ps2} we plot the normalized mass function, setting
$S_1 = \sigma^2(M_1)$, $S_2 = \sigma^2(M_2)$, $\xi =
\xi_{r_{max}}(d,M_1,M_2)$, and fixing $\nu_1 = \nu_2 = \delta_c(z)$
for various redshifts, in the same $\Lambda$CDM cosmology considered
in \S3.  Each function is normalized by its uncorrelated value in
order to emphasize the features of the joint distribution, i.e., each
panel shows $\frac{d^2 n^2_{12}}{d M_1 d M_2} \left[\frac{d n_1}{d
M_1}\times \frac{d n_2}{d M_2} \right]^{-1}$ as a function of $M_2$,
for various values of $d$, $M_1$, and $z$.

In the left panels of this figure we fix $M_1$ to be $10^{12} \hMsun$,
corresponding to a typical $L_\star$ galaxy, and construct the
normalized number density at redshifts $1$ and $2$, and Lagrangian
distances of $d=1$, 3.3, and 10 comoving $\hMpc$. At the larger
distances, the presence of a perturbation at point $A$ enhances the
formation of an object at point $B$, unless $M_2$ is so large that
such a halo at point $B$ would very likely absorb point $A$ into it as
well. Thus the $d$ = 3.3 and 10 comoving $\hMpc$ curves are enhanced
at all but the largest $M_2$ values.  This enhancement is more
significant at $z=2$ as these peaks are rarer, and hence more highly
biased.

As the points get closer, however ($d =1 \hMpc$), the lines become
strongly peaked at $M_2 = M_1$, excluding the formation of objects of
vastly different sizes at a short distance. Indeed, as $d \rightarrow
0$, $f$ approaches the single point differential mass function times a
delta function, as both points must belong to the same peak in the
limit when the two points become identical. Real applications of these
results at short distances must consider more explicitly the issue of
halo exclusion, i.e., the fact that a given halo at point $A$ contains
all the mass from some region, and this halo either contains the point
$B$ or not. More generally, the two regimes $M_2 < M_1$ and $M_2 >
M_1$ should be considered separately because of their different
physical interpretations (especially when $d$ is small); in the first
case, $M_2$ is a small halo that may be accreted by $M_1$, and in the
second case, $M_2$ is a large halo that may be about to absorb $M_1$
into it.

In the right panels, we consider the case of a dwarf galaxy with $M_1
= 10^9 \hMsun$, forming at redshifts $z=4$ and $z=6$, such that
$\nu(z)/\sigma(M_1)$ and $\nu(z)/\sigma(M_2)$ cover a similar range of
values as in the $L_\star$-galaxy case.  As the virial radii of these
systems are smaller by a factor of 10, in these panels we fix $d$ =
0.1, 0.33, and 1.0 comoving $\hMpc$. These plots display essentially
the same features as in the more massive case, although the effects of
the correlations are somewhat larger since the objects are slightly 
rarer. 

\begin{figure}
\plotone{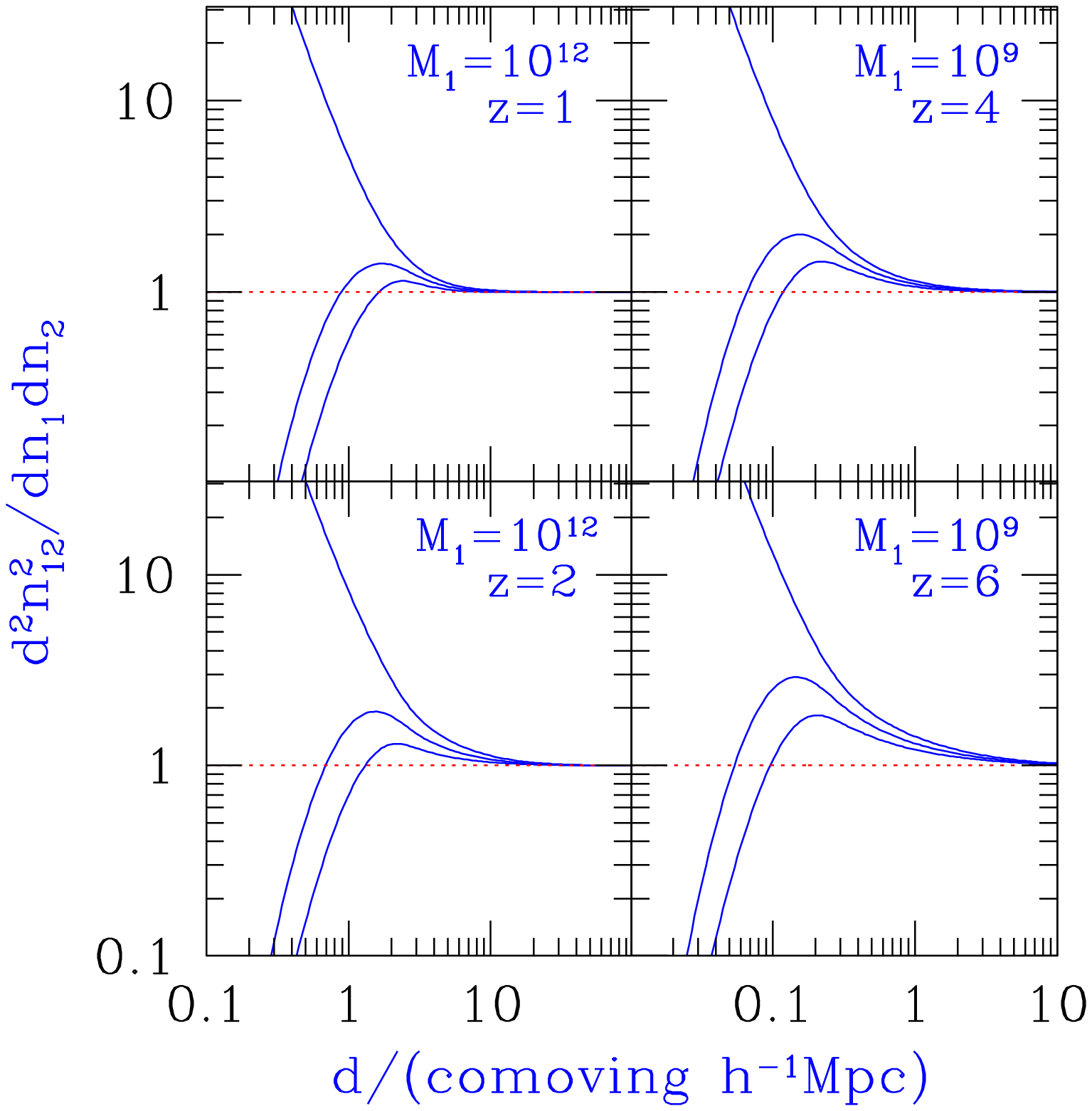}
\caption{Normalized number density of correlated halos as calculated
from eq.\ (\ref{eq:2ptm}) as a function of $d$, for various values of
$M_1$, $M_2$, and redshift. Again in the left panels $M_1$ is held
fixed at a value of $10^{12} \hMsun$ at redshifts of 1 and 2, while
the solid lines show, from top to bottom (in terms of the highest
point in each curve), densities with $M_2$ fixed at $10^{12} \hMsun$,
$3.3 \times 10^{11} \hMsun$ and $10^{11} \hMsun$, respectively, and
again the uncorrelated case is plotted as a dotted line at
1. Similarly, in the right panels, we fix $M_1$ to be $10^9 \hMsun$ at
redshifts 4 and 6.  In these panels the solid lines show normalized
densities at $M_2$ = $10^9 \hMsun$, $3.3
\times 10^8 \hMsun$, and $10^8 \hMsun$, from top to bottom.}
\label{fig:ps2d}
\end{figure}

In Figure \ref{fig:ps2d} we again plot the normalized number density
of two points forming at the same redshift, but now holding $M_2$
fixed and varying the distance between the two perturbations.  Here we
see that as the points come closer to each other, the number densities
are initially enhanced, as the collapse of an overdensity at the first
point makes it likely that a large-scale overdensity enhances halo
formation at all nearby points. As the perturbations are drawn closer
together, the curves drop sharply if $M_1 \neq M_2$ as it is
impossible for two perturbations of different sizes to form at the
same position and redshift. If $M_1 = M_2$ however, the probability
continues to rise dramatically at small distances, as $f$ approaches
the single point differential mass function multiplied by a delta
function. Again these effects are stronger at higher redshifts, as
rare peaks are more highly correlated.

\begin{figure}
\plotone{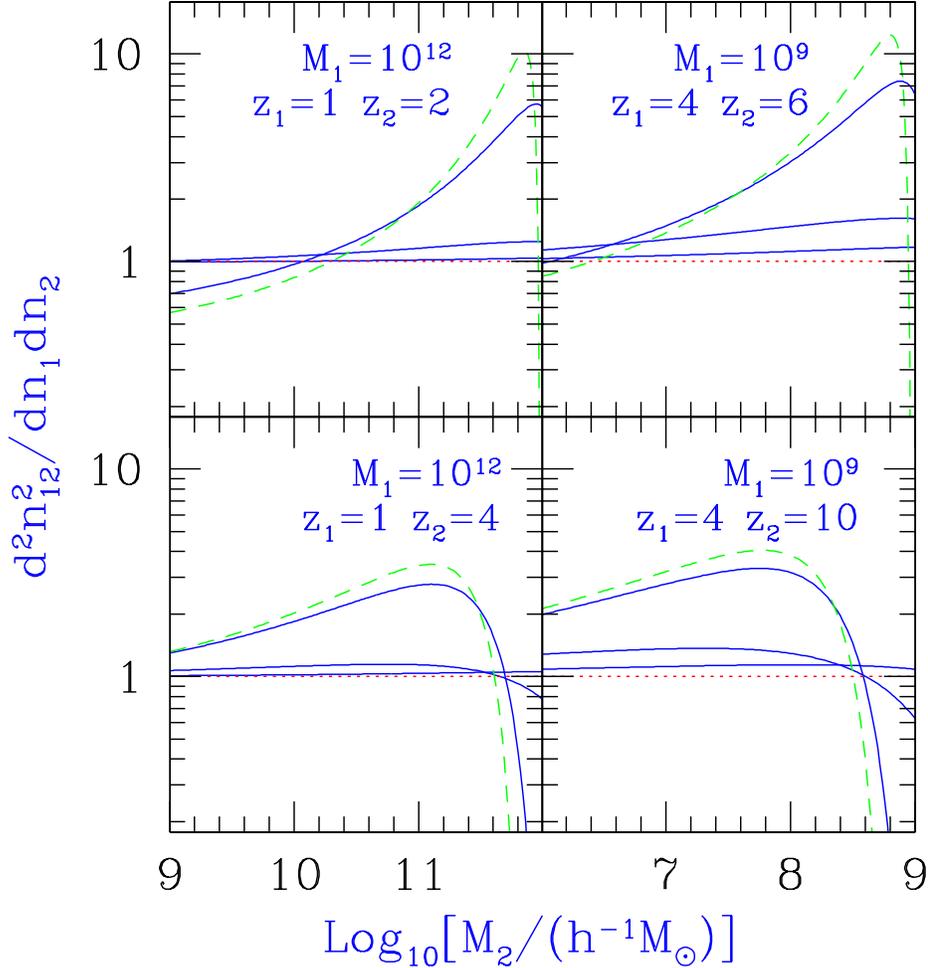}
\caption{Normalized number density of halos as a function of $M_2$ for
two objects forming at different redshifts. In the left panels, we fix
$M_1 = 10^{12} \hMsun$ and $z_1=1$, and consider the second object as
a generalized progenitor, forming at a higher redshift at a point
nearby.  In these panels, each of the solid lines correspond, from top
to bottom (in terms of the highest point in each curve), to $d$ values
$1$, $3.3$, and $10$ comoving $\hMpc$, and we consider cases in which
$z_2 =2$ and $z_2 = 4.$ In the right panels, we fix $M_1 = 10^{9}
\hMsun$ and $z_1=4$, while the solid lines now correspond, from top to
bottom, to $d$ values of $0.1$, $0.33$, and $1.0$ comoving $\hMpc$,
with $z_2= 6$ and $z_2 = 10.$ Also shown for reference in all panels
are the dotted lines at $1$, corresponding to the uncorrelated case,
and the fully-correlated dashed lines, corresponding to the formation
of both objects at the same point, as given by eq.\
(\protect\ref{eq:merger}).  }
\label{fig:ps22}
\end{figure}

\begin{figure}
\plotone{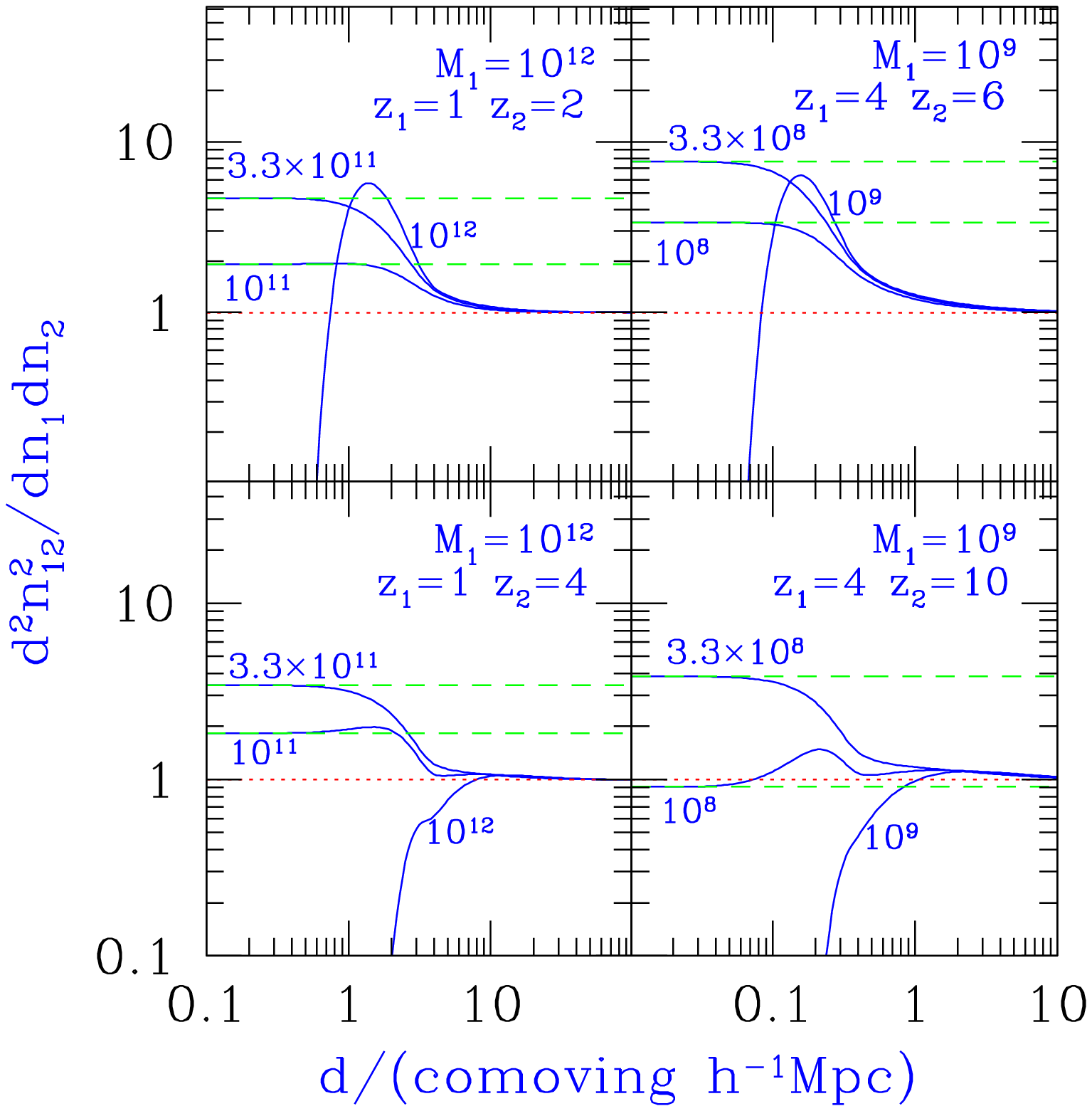}
\caption{Normalized number density of halos as a function of
$d$ for two objects forming at different redshifts.  In the left
panels we fix $M_1 = 10^{12} \hMsun$ and $z_1=1$, and consider $M_2$
values of $10^{11}$, $3.3 \times 10^{11}$, and $10^{12} \hMsun$, at
redshifts $z_2 =2$ and $z_2 = 4.$ In the right panels, we consider a
smaller, higher redshift object and fix $M_1 = 10^{9} \hMsun$
and $z_4=4$. In this case we take $M_2$ values of $10^{8}$, $3.3
\times 10^{8}$, and $10^{9} \hMsun$, at redshifts $z_2 =6$ and $z_2 =
10.$ Each solid line is labeled by its corresponding $M_2$ value,
while the fully uncorrelated and fully-correlated cases are shown in
all cases by the dotted and dashed lines respectively (except that for
$M_1=M_2$, the fully-correlated limit is a normalized density of
zero).}
\label{fig:ps2d2}
\end{figure}

Our formalism is not restricted to the collapse of two points at the
same redshift, but is also well suited to study the formation of halos
at varying times.  In Figures \ref{fig:ps22} and \ref{fig:ps2d2} we
again examine the formation at $z=1$ of an $M_1 = 10^{12} \hMsun$ halo
associated with an $L_\star$ galaxy, and the formation at $z_1 = 4$ of
an $M_1 = 10^{9} \hMsun$ halo associated with a dwarf galaxy. We now
consider $M_2$ to be a smaller halo, formed at a distance $d$ and at
an earlier redshift.  This is essentially a generalization of the
usual progenitor problem, where now the lower mass halo need not be
absorbed into the larger halo, but can be located an arbitrary
distance away.

In Figure \ref{fig:ps22} we consider the normalized number density as
a function of $M_2$ for a variety of distances, fixing $z_1 = 1$ and
$z_2 = 2 $ and 4 in the $M_1 = 10^{12} \hMsun$ case shown in the left
panels, and fixing $z_1 = 4$ and $z_2 = 6$ and 10 in the $M_1 = 10^{9}
\hMsun$ case shown in the right panels. Many features in this
plot parallel the $z_1 = z_2$ cases: at the larger distances, the
curves are enhanced at most $M_2$ values, as a second galaxy is likely
to form near the first; and if the objects are too close together and
have masses that are very different then they can suppress each
other's likelihood of formation.

Unlike the simultaneously collapsing cases, when $M_1 = M_2$, $f$ does
not approach a delta function as $d \rightarrow 0$. Instead, the
formation of objects of equal mass at both points is completely
excluded, as this would correspond to the same halo forming in the
same place at two different times (while the formalism assumes
continuous mass accretion for every halo). 
Each curve is instead peaked at an $M_2$ value that is {\em
smaller}\, than $M_1$, as this value roughly corresponds  to the most
likely progenitor at a redshift $z_2$ for the larger halo, which forms
at $z_1$. These curves can be compared to the number densities
expected for a progenitor at the same point (i.e., at $d=0$), first
derived by
\citet{bc91},
\be
\frac{\nu_2-\nu_1 }
{\sqrt{2 \pi} (S_2-S_1)^{3/2}}
\exp\left[-\frac{(\nu_2-\nu_1)^2}{2 (S_2-S_1)}\right]
\frac{\nu_1 }
{\sqrt{2 \pi} S_1^{3/2}}
\exp\left[-\frac{\nu_1^2}{2 S_1}\right],
\label{eq:merger}
\ee
which is shown by the dashed lines.  Note that as $d \rightarrow 0$,
$f$ approaches this expression exactly.

In Figure \ref{fig:ps2d2} we study the generalized progenitor problem
as a function of distance, again fixing $z_1 = 1$ and $z_2 = 2 $ and 4
when $M_1 = 10^{12} \hMsun$, and fixing $z_1 = 4$ and $z_2 = 6$ and 10
when $M_1 = 10^{9} \hMsun$.  In the upper panels, in which the
differences in redshifts are relatively small, the normalized
bivariate number density is in general more strongly enhanced, the
larger the progenitor object and the closer it lies to the lower
redshift halo $M_1.$ The only limit in which this enhancement does not
increase as the objects get closer, in fact, is the limit in which
$M_2 \rightarrow M_1$, which must be excluded as the same object
cannot form at the same point twice.

As the difference between $z_1$ and $z_2$ becomes greater, however,
the behavior of the normalized number density becomes more complex, as
is illustrated in the lower panels. We find in this case that at all
except the largest distances, $\frac{d^2 n^2_{12}}{d M_1 d M_2}
\left[\frac{d n_1}{d M_1}\times \frac{d n_2}{d M_2} \right]^{-1}$ no
longer increases monotonically as a function of $M_2$. This is because
even when $d$ is large enough that it is physically possible to form
objects with masses $M_2 = M_1$ at both points, the presence of a
large object at high redshift at point $B$ makes it likely that point
$A$ will instead be absorbed into an {\em even larger}\, halo that
forms at point $B$ at some intermediate redshift $z_1 < z < z_2.$
Thus, the case $M_1 = M_2$ is suppressed for almost all $d$ values,
while the densities are actually enhanced in the presence of a more
modest high-redshift perturbation. Note that again in all cases $f$
approaches eq.\ (\ref{eq:merger}) exactly at small distances,
reproducing the usual progenitor distribution at a single point.

\subsection{Bivariate Cumulative Mass Fraction and Nonlinear Bias}

Returning to eq.~(\ref{eq:bigF}), we can construct
the fraction of trajectories that have been absorbed
by both barriers. Writing $Q$ in terms of the $x$ integral
and performing the appropriate $\delta_1$ and $\delta_2$ integrals
this becomes
\be
F(\nu_1,\nu_2,S_1,S_2,\xi) = \int_{-\infty}^{\nu_{\rm min}} dx\,
\left[ G(x, \xi) - G(2 \nu_{\rm min} - x, \xi) \right]\,
{\rm erf} \left( \frac{\nu_1-x} {\sqrt{2 (S_1-\xi)}} \right)
	{\rm erf} \left( \frac{\nu_2-x} {\sqrt{2 (S_2-\xi)}} \right)\ .
\label{eq:Fbias}
\ee
This is the product of the mass fraction in halos of mass below
$M_1(S_1)$ at redshift $z_1(\nu_1)$ and the mass fraction in halos of
mass below $M_2(S_2)$ at $z_2(\nu_2)$, given that the distance between
the halos is $d$. In other words, if we divide $F$ by the one-point
value of the mass fraction in halos of mass $<M_1(S_1)$ at
$z_1(\nu_1)$, we obtain the biased mass fraction in halos of mass
$<M_2(S_2)$ at $z_2(\nu_2)$, where the biasing refers to an average
only over points at a distance $d$ from the population of halos
$<M_1(S_1)$. In order to construct the correlation function of rare,
massive halos, we instead need a closely related quantity; this is the
``bivariate cumulative mass fraction,'' that is the product of mass
fractions in halos with masses {\em above}\, $M_1$ and $M_2$, given by
\be
F(\nu_1,\nu_2,<S_1,<S_2,\xi) = 1 + F(\nu_1,\nu_2,S_1,S_2,\xi) - {\rm
erf} \left( \frac{\nu_1} {\sqrt{2 S_1}}\right) - {\rm erf} \left(
\frac{\nu_2} {\sqrt{2 S_2}}\right)\ .
\label{eq:Flessthan}
\ee

\begin{figure}
\plotone{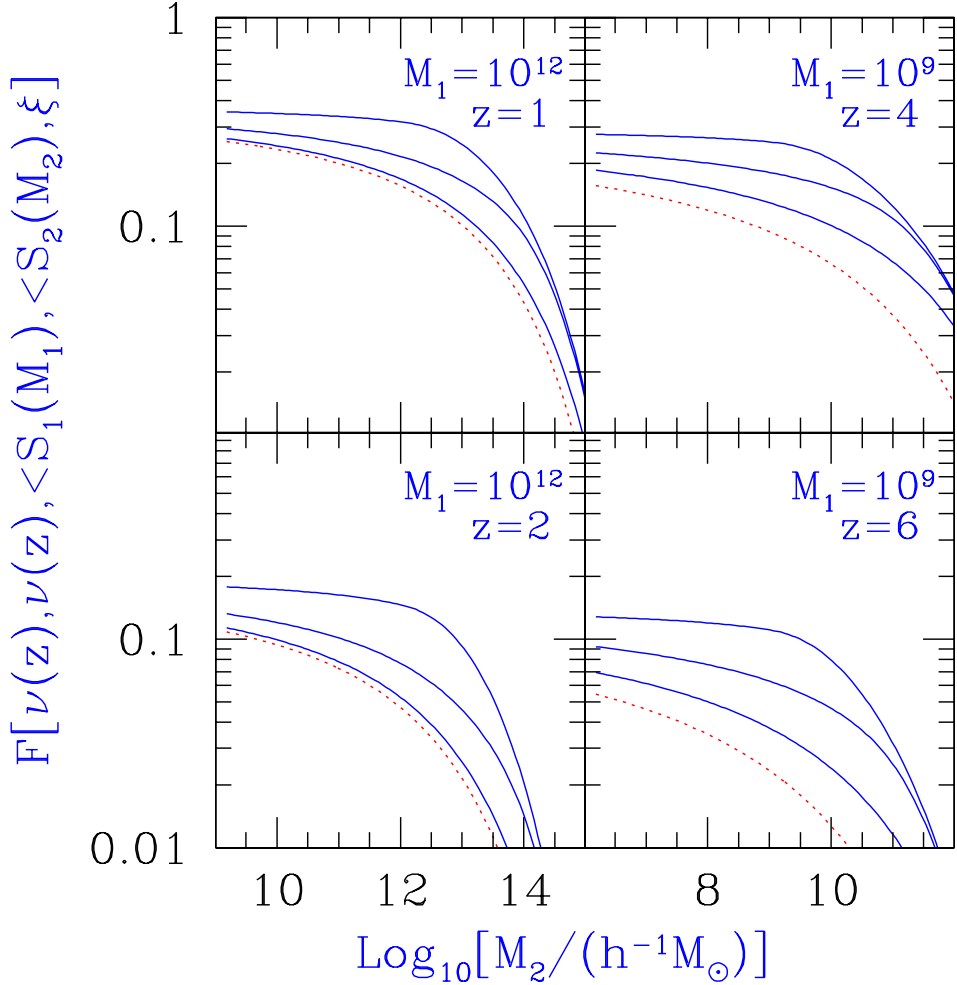}
\caption{Bivariate cumulative mass fraction as a function of
$M_2$, for various values of $d$, $M_1$, and $z$. In the left panels
$M_1=10^{12} \hMsun$, and $z = $ 1 and 2. From top to bottom, the
solid lines in these panels show $F$ values with the Lagrangian
distances between the two points fixed at 1.0, 3.3, and 10 comoving
$\hMpc$ respectively, and the uncorrelated case is shown for reference
by the dotted lines.  In the right panels, $M_1=10^9 \hMsun$, and $z
=$ 4 and 6.  In these panels the solid lines show, from top to bottom,
$F$ values with d = 0.1, 0.33, and 1.0 comoving $\hMpc$, while the
dotted lines again correspond to the uncorrelated case.}
\label{fig:Fm}
\end{figure}

In Figure \ref{fig:Fm} we plot $F(\nu_1,\nu_2,<S_1,<S_2,\xi)$, setting
$S_1 = \sigma^2(M_1)$, $S_2 = \sigma^2(M_2)$, and $\xi =
\xi_{r_{max}}(d)$, and fixing $\nu_1 = \nu_2 = \delta_c(z)$ at various
redshifts. In the left panels of this figure we again set $M_1 =
10^{12} \hMsun$, $z = 1$ and $2$, and $d=1$, 3.3, and 10 comoving
$\hMpc$. At the smallest distances, the two points are almost
completely correlated, and $F$ approaches its one-point value of ${\rm
erfc} [\nu(z)/\sqrt{2 \sigma^2(M_{\rm max})}]$ as given by eq.\
(\ref{eq:fm1point}), where $M_{\rm max}$ is the greater of the two
masses.  This is because the plotted quantity expresses the
joint probability of a halo of mass $>M_1$ at point $A$ and a halo
$>M_2$ at point $B$. When the two points are very close together, both
points are contained within the same halo, which therefore must have a
mass greater than $M_{\rm max}$. Thus if, e.g., $d = 1 \hMpc$ and $M_2
\leq M_1$, $F$ is roughly constant at a value corresponding to $M_1$,
while $F$ quickly approaches ${\rm erfc} [\nu(z)/\sqrt{2
\sigma^2(M_2)}]$ when $d = 1$ and $M_2 > M_1$. Note also that at $d =
3.3 \hMpc$ the points are somewhat uncorrelated at small values of
$M_2$, but at larger value of $M_2$ the overlap between the points
becomes much larger, and thus they become almost completely
correlated, moving towards the $d = 1 \hMpc$ case at the highest $M_2$
values. Finally, at $d=10 \hMpc$, the points are almost uncorrelated
and $F$ closely approximates the product of two independent one-point
probabilities, as given by the dotted lines. At this large distance,
correlations are only somewhat significant at the highest $M_2$
values, moving $F$ slightly away from the dotted lines.

In the right panels, we again consider $M_1 = 10^9 \hMsun$, $z=4$ and
$z=6$, and $d$ = 0.1, 0.33, and 1.0 comoving $\hMpc$. The main
features are similar to those seen in the higher-mass cases, although
the correlations between the two points are somewhat stronger, moving
the $d$ = 1.0 and 0.33 curves away from the dotted, uncorrelated cases,
and up towards the almost fully correlated $d$ = 0.1 line. Again this
line is roughly constant at a value corresponding to $M_1$ for low
values of $M_2$, and moves towards ${\rm erfc} [(\nu(z)/\sqrt{2
\sigma^2(M_2)}]$ at larger $M_2$ values, and again the correlations
for the $d$ = 0.33 and 1.0 $\hMpc$ cases become more significant at
large values of $M_2.$

Returning to eq.\ (\ref{eq:Flessthan}), we can also immediately obtain
the biasing of rare halos; namely, the bias between halos of masses
larger than $M(S)$ at a distance $d$ at a redshift $z(\nu)$, defined
as the increase in the cumulative mass fraction at the second point
given the presence of an object at the first point. This is simply
given by the bivariate cumulative mass fraction divided by the
cumulative mass fraction at two uncorrelated points,
\be
\xi_{\nu,{\rm 2stp}}(d,S) + 1 =
\left[ {\rm erfc}\left({\frac{\nu}{\sqrt{2 S}}}\right) \right]^{-2}
F(\nu,\nu,<S,<S,\xi(d,S))\ .
\label{eq:ourxinu}
\ee

This equation can be compared with the usual expression used for
nonlinear bias \citep{kaiser84}, which we can rewrite in a form
similar to eq.\ (\ref{eq:Fbias}) as
\be
\xi_{\nu,\rm K84}(d,S) + 1 =
\left[{\rm erfc}\left({\frac{\nu}{\sqrt{2 S}}}\right)\right]^{-2}
\int_{-\infty}^{\infty} dx\,
G(x, \xi) \,
\left[
{\rm erfc} \left( \frac{\nu - x} {\sqrt{2 (S-\xi)}} \right) \right]^2.
\label{eq:k84}
\ee
This expression was derived by simply integrating over the
distribution of probabilities at each of the two points, in a manner
analogous to Press \& Schechter's original derivation of the
one-point mass function. Thus, this expression suffers from the
same peaks-within-peaks problem which forced Press \& Schechter to
multiply their expressions by an arbitrary factor of 2.

\begin{figure}
\plotone{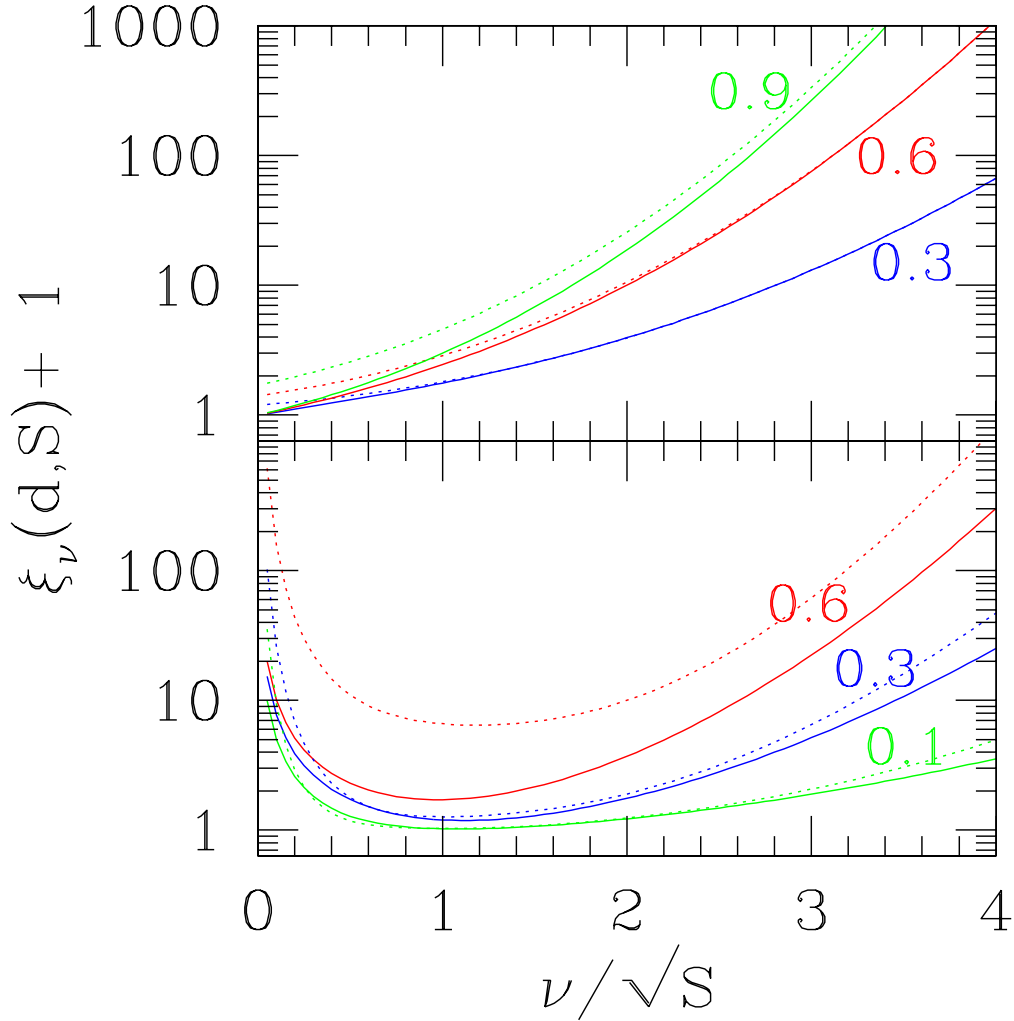}
\caption{
Nonlinear bias of rare peaks, as a function of $\nu/\sqrt{S}$. The
cumulative bias is shown in the top panel; in each pair of curves the
dotted line corresponds to the standard expressions derived in
\protect\citet{kaiser84}, while the (lower) solid line correspond to
$\xi_{\nu,{\rm 2stp}}.$ Each pair of lines is labeled in terms of the
scaled correlation between the two points, $\xi/S$, which is 0.9, 0.6,
and 0.3, from top to bottom. The bias of halos at a given mass is
shown in the bottom panel; in each pair of curves the dotted line
shows the biased correlation function that is based on
\protect\citet{mo96}, while the solid line corresponds to
$\hat{\xi}_{\nu,{\rm 2stp}}$. In this case $\xi/S$ values of 0.6, 0.3,
and 0.1 are used, from top to bottom.}
\label{fig:bias}
\end{figure}

The bias between peaks as computed using the two-step approximation,
eq.~(\ref{eq:ourxinu}), and the one from the standard approach,
eq.~(\ref{eq:k84}), are plotted in Figure~\ref{fig:bias} as a function
of the height of the peak, in units of $\sqrt{S}$, for two
perturbations of various correlation strengths $\xi/S$.
This comparison helps clarify the range over which eq.\ (\ref{eq:k84})
is most accurate, but it also makes clear that eq.\ (\ref{eq:ourxinu})
is a superior generalization which self-consistently accounts for all
of the matter in the universe.

In the fully-correlated limit ($\xi \rightarrow S$), $\xi_{\nu,K84}+1$
is too high by a factor of 2, at all $\nu$. This is because in this
case, the points are fully correlated so the joint fraction is
$F_{1,2} = F_1$, where $F_1={\rm erfc}(\nu/\sqrt{2 S})$ is the
cumulative mass fraction of a single halo, and therefore the bias is
given simply by $F_{1,2}/F_1^2 = 1/F_1$. While the two-step derivation
leads to the correct bias in this limit, Kaiser's derivation ignores
the Press-Schechter factor of 2 caused by the problem of peaks within
peaks, and thus $\xi_{\nu,K84}+1$ approaches $2/F_1.$ Note also that
in the two-step case, for all values of $\xi/S$, at $\nu/\sigma
\rightarrow 0$ the bias approaches 1. This is because every point is
in a halo corresponding to some $\nu/\sigma > 0$, and thus imposing
$\nu/\sigma > 0$ is no constraint at all and does not result in
bias. In the Kaiser (1984) case, however, this mass conservation is
not imposed and instead $\xi_{\nu,\rm K84}\rightarrow \xi/S$ as
$\nu/\sigma \rightarrow 0$. Finally, at large values of $\nu$, when
$\nu \gg \sigma/(1 - \xi/S)$, the two expressions become equal. Note,
however, that for sufficiently high values of $\xi/S$, $\xi_{\nu,{\rm
K84}}$ suffers from the erroneous factor of 2 problem even in the case
of extremely rare peaks.

Finally, we consider another definition of bias which is also commonly
used. This definition is designed to yield more directly the
(Lagrangian) correlation function among rare peaks at a {\it given}\,
mass $M$, rather than the bias among the cumulative halo population
above some mass $M$. In our calculation, this correlation function is
simply given by
\be
\hat{\xi}_{\nu,{\rm 2stp}}(d,S) + 1 =
\left[ f(\nu,S) \right]^{-2} f(\nu,\nu,S,S,\xi(d,S))\ ,
\label{eq:ourxinu2}
\ee
where we refer to eqs.~(\ref{eq:f1pt}) and (\ref{eq:littlef}). In the
absence of any previous derivation of this correlation function, even
in the simple case corresponding to the assumptions of Kaiser (1984),
previous efforts have resorted to defining the bias in other ways and
hoping that these alternative definitions yield a value which is close
to the desired correlation function. These methods include the
``peak-background split'', which gives \citep{ck89}
\be
\hat{\xi}_{\nu,{\rm pk-bgd}}(d,S) = \frac{\xi}{S}\, \left( \frac{\nu^2/S 
- 1} {\delta_c} \right)^2\ .
\label{eq:pk-bgd}
\ee In testing these alternative definitions we assume that the ratio
between the halo correlation function ($\hat{\xi}_{\nu,{\rm pk-bgd}}$
in this case) and the normalized mass correlation function ($\xi/S$)
equals the square of the bias. Eq.\ (\ref{eq:pk-bgd}) is derived only
in the limit of rare halos and large $d$, so we consider its
generalization \citep{mo96}, \be \hat{\xi}_{\nu,{\rm MW96}}(d,S) =
\frac{\xi}{S}\ \frac{\int_{-\infty}^{1} Q(\nu, \nu x, S) \left[ \frac{
f(\nu (1-x),S-S_0)} {f(\nu,S)} -1 \right]^2 dx} {\int_{-\infty}^{1}
Q(\nu, \nu x, S)\, \del_c^2\, x^2\, dx}\ .
\label{eq:mo96}
\ee In this expression, $\del_c=1.686$ (see \S 2) and $S_0$ is the
$\sigma^2$ of the mass scale $M_0$, where a sphere (at the mean
density) of comoving radius $d$ contains a mass $M_0$. Note that
$\hat{\xi}_{\nu,{\rm MW96}}$ was calculated from
eq.~(\ref{eq:merger}), using a definition of bias which considers the
number density of halos of a given mass that form out of matter
initially contained within larger overdense spheres. The comparison in
Figure~\ref{fig:bias} (bottom panel) shows that $\hat{\xi}_{\nu,{\rm
MW96}}$ successfully approximates $\hat{\xi}_{\nu,{\rm 2stp}}$ only
when $\xi/S$ is small, and only over a limited range of values of
$\nu/\sigma$. On the other hand, our result represents a direct
analysis of the correlation function of halos, and at all parameter
values it is fully consistent with the closely related bias
$\xi_{\nu,{\rm 2stp}}$ of eq.~(\ref{eq:ourxinu}).

The abundance of halos and their correlations are determined by the
linear power spectrum, since the formation of a halo is a long-term
process that is driven by the existence of initial overdensities.
However, correlation functions involving parts of halos depend also on
the non-linear, internal structure of the halos themselves. Examples
include the autocorrelation function of galaxies, which depends on the
variation with halo mass of the number of galaxies per halo, and the
dark matter autocorrelation function, which depends on the internal
density profiles of halos. Prescriptions for the internal structure of
halos have been previously combined with the halo correlation function
based on \citet{mo96} to produce various autocorrelation functions
\citep[e.g.,][]{s00,mf00}. These analyses can now be improved with
our generalized, self-consistent results for halo correlations.

\subsection{Bivariate Mixed-mass Function}

The last quantity of interest is the bivariate mixed mass function,
which is the mass function at one point times the correlated 
cumulative mass fraction at a second point,
\be
\frac{dn_1 F_2}{dM_1} =
\frac{\bar{\rho}}{M_1} \left|\frac{d S_1}{d M_1} \right|
{\cal F}(\nu_1,\nu_2,S_1,S_2,\xi)\ .
\ee
Using eqs.~(\ref{eq:f1}) and (\ref{eq:dxQ0}), this can be written 
as the sum of three terms:
\ba
{\cal F}  &=& 	4 \frac{\partial \xi} {\partial S_1} Q_0 +
\frac{\nu_1}{S_1} G(\nu_1, S_1)\nonumber \\
&- &\int_{-\infty}^{\nu_{\rm min}} dx\,
	\left[ G(x, \xi) - G(2 \nu_{\rm min} - x, \xi) \right]\,
	\frac{\nu_1-x} {S_1-\xi}\, G(\nu_1-x, S_1-\xi)\, {\rm erf} \left(
	\frac{\nu_2-x} {\sqrt{2 (S_2-\xi)}} \right), \label{Mixedint}
\label{eq:mixed}
\ea
where $Q_0$ is
evaluated at $\delta_1=\nu_1$ and $\delta_2=\nu_2$.

\begin{figure}
\plotone{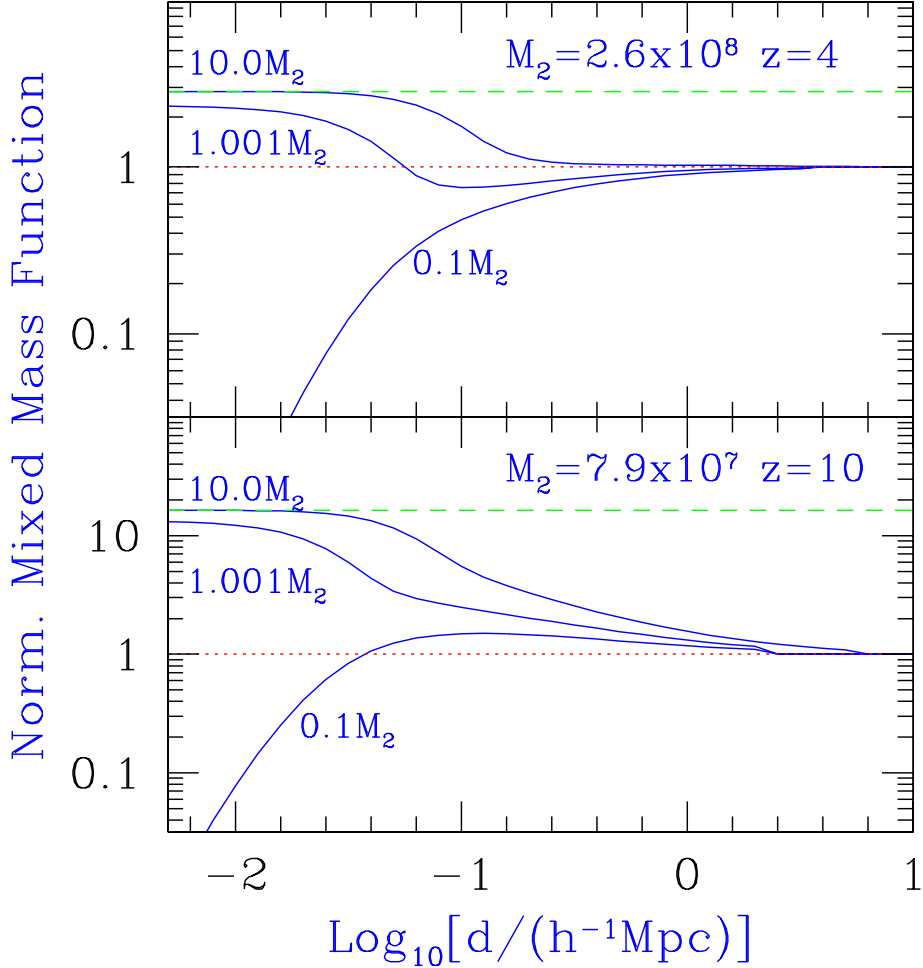}
\caption{Bivariate Mixed-mass function.  {\em Upper panel:}\, each
sold line shows ${\cal F}$ normalized by its uncorrelated value, as a
function of $d$, for various values of $M_1$.  The lines are labeled by
their $M_1$ values and in all cases $z_1 = z_2 = 4$, and $M_2$ is the
cooling mass at this redshift, $2.6 \times 10^8 M_\odot$.  For
reference, the fully uncorrelated line is given by the dotted line and
the fully correlated case in which $M_1 > M_2$ is shown by the dashed
line; if $M_1 < M_2$ then the fully correlated case is zero. {\em
Lower panel:}\, Normalized ${\cal F}$ for a higher redshift case in
which $z_1 = z_2 = 10$, and $M_2$ is the (smaller) cooling mass at
this redshift, $7.9 \times 10^7 M_\odot$, which corresponds to
slightly rarer objects. The lines follow the same conventions as in
the upper panel.}
\label{fig:mix}
\end{figure}

In Figure \ref{fig:mix} we plot this quantity, again normalizing by
its limiting value at large distances, \be {\rm erfc}
\left(\frac{\nu_2}{\sqrt{2 S_2}} \right) \frac{\nu_1}{\sqrt{2 \pi}
S_1^{3/2}} \exp \left(-\frac{\nu_1^2}{2 S_1} \right)\ , \ee in order
to emphasize its overall features.  The plotted quantity equals the
one-point mass function $f(\nu_1,S_1)$ of $M_1$ halos times the total
mass fraction, at a distance $d$ away, in halos above mass $M_2$. In
this plot we restrict our attention to halos that are able to cool in
the neighborhood of the halo of mass $M_1$; for illustration we assume
that molecular hydrogen formation is inefficient, and that star
formation occurs only in galaxies in which atomic cooling is
efficient. This in turn requires a halo virial temperature of at least
$10^4$ K, which sets a minimum value of $M_2$ at each redshift.

The figure shows that if $M_1$ is much smaller than $M_2$, then at
small distances, the mass fraction above the cooling limit is zero, as
no objects larger than $M_1$ form at this point.  At intermediate
values of the distance, however, the presence of a small object is
able to either lower or raise the cooled mass fraction depending on
how common or rare the objects are.

If $M_2$ is a generalized progenitor of $M_1$, however, so that $M_1 >
M_2$ and $z_2 \geq z_1$, then we can compare ${\cal F}$ with the mixed
mass function for two fully correlated points,
\be 
{ \rm erfc} \left[\frac{\nu_2-\nu_1 }{\sqrt{2 (S_2-S_1)}} \right]
\frac{\nu_1}{\sqrt{2 \pi} S_1^{3/2}} \exp\left(-\frac{\nu_1^2}{2
S_1}\right)\ ,
\label{eq:mixedc}
\ee
which corresponds to the dashed lines in the figure. In the case in
which the two objects form at the same redshift, this is simply the
mass function at a single point, as at short distances the presence of
an object of a size larger than the cooling mass indicates that all
the gas at this point has cooled. In general, the presence of a halo
with $M_1 > M_2$ can produce either a positive or negative overall
bias in the amount of gas that cools within nearby halos, depending on
$d$ and on $M_1$.
  
\section{Conclusions}

In this work, we have developed a simple but powerful extension of the
\citet{ps74} model that goes beyond \citet{bc91} and can be used to
study inhomogeneous structure formation and correlation functions
among halos. Although the linear excursion-set model represents the
backbone of our modern analytical understanding of structure
formation, it has previously only been able to predict the average
number density of collapsed objects and has not supplied any
information as to their relative positions. While many authors have
derived approximate expressions for spatial correlations either
analytically or from numerical simulations, each of these was grafted
externally onto the underling formalism, leaving inhomogeneous
structure formation on slippery theoretical ground.

Although a strict excursion set description of inhomogeneous structure
formation can be constructed only by solving the diffusion equation
numerically, we have shown in this work that such solutions can be
matched to within $2\%$ accuracy using a simple approximation. By
considering the trajectories of the overdensities at two points to be
at first fully correlated and later fully uncorrelated as a function
of decreasing filter scale, and by carefully choosing the cutoff value
between these two regimes, we have developed an analytical formalism
appropriate for studying problems in which a simple one-point approach
is insufficient. 

With this two-step approximation we are able to derive an
approximation to the joint probability distribution for the
overdensities of two points with arbitrary mass scales and collapse
redshifts, separated by any comoving Lagrangian distance. This
distribution leads directly to an analytical expression for the
bivariate mass function of halos, given by eqs.~(\ref{eq:littlef}) and
(\ref{eq:2ptm}), and for the bivariate cumulative mass fraction, given
by eqs.~(\ref{eq:Fbias}) and (\ref{eq:Flessthan}). These expressions
generalize a number of previous results, and also yield
self-consistent expressions for the nonlinear biasing and correlation
function of halos, given by eqs.~(\ref{eq:ourxinu}) and
(\ref{eq:ourxinu2}).  Our results also incorporate halo exclusion,
i.e., the fact that if point $A$ belongs to a given halo, all nearby
points $B$ are likely to belong to the same halo (see, e.g.,
Figure~\ref{fig:ps2d}).  We have also shown that the two-step
approximation yields the right value in every limit where it must
match a result derived from the one-point approach. We have provided
{\em Gemini}\, (see Appendix B), a publicly-available code that makes
it easy to apply our formalism in specific cases.

Our results form an analytical framework that can be used in
conjunction with numerical simulations to study inhomogeneous
structure formation in a manner similar to the approach commonly
adopted to study average quantities; analytical techniques are used to
outline the overall physical picture and quantify model uncertainties,
and numerical techniques are used to refine these results through a
limited series of detailed tests. Likewise, these comparisons must
take into account many of the issues that arise in a spatially
averaged context, such as the best choice of collapse density
$\delta_c$ \citep[e.g.,][]{kt96,smt01,jenk01}, possible corrections to
the overall functional form of the mass function
\citep[e.g.,][]{jenk01,ls98}, and the relationship between the
Eulerian coordinate system which is observed and the Lagrangian
coordinates that are used in the excursion-set formalism
\citep[e.g.,][]{mo96,cat98,j99}. While the study of such issues will
sharpen the link between this formalism and more directly observable
quantities, our method nevertheless already provides an important
first step towards a better theoretical understanding of inhomogeneous
structure formation.  In this context the Lagrangian coordinate system
intrinsic to an excursion-set description is a mixed blessing, for
although it is more difficult to compare with simulations, it is
usually much more important physically to have a measure of the total
column depth of material separating two objects than their precise
distance.

For many years the study of structure formation has centered on the
formation of individual dark matter halos and their direct
progenitors, yet many classes of problems that are now being studied
can not be addressed in this context; and while it was once assumed
that gravity alone controlled cosmic evolution, many more recent
issues in structure formation are better described as an interplay
between the IGM and the objects that form within it
\citep[e.g.,][]{bl99,cf00,st01}. Each new generation of objects
changes the state of the gas, and this state in turn affects the
properties of the next generation to form.

>From the dissociation of molecular hydrogen, to reionization and the
resulting photoevaporation of halos, to the epoch of galactic winds,
our universe has time after time undergone intense and inhomogeneous
transformations.  Although the intricacies of such processes will
undoubtedly take years of observational and theoretical investigation
to unravel, it is clear already that one's answers will depend on
one's relative location.

\acknowledgments

We thank Dick Bond, Yuri Levin, Avi Loeb, and Anatoly Spitkovsky for
useful comments and discussions.  ES has been supported in part by an
NSF MPS-DRF fellowship.  RB acknowledges support from CITA.

\section*{Erratum}

Immediately preceding the publication of this article we were alerted
to the existence of a previous investigation into inhomogeneous
structure formation from the point of view of excursion sets. In
Porciani \etal (1998), the authors considered the simultaneous
evolution of two points separated by a fixed initial comoving
distance, but rather than approximate this evolution with a two-step
procedure such as the one described in \S 3.3 of Scannapieco \&
Barkana (2002) they instead made use of the correlated two-point
solution {\em in the absence of barriers}\/ and imposed on this the
same reflecting boundary conditions as in the fully uncorrelated case.
In this way they were able to develop what amounts to an alternate
approximation to our eq.\ (41) in the limited case of simultaneously
collapsing halos.

While both approaches yield similar results at small values of
$\xi/S$, the approximation proposed by Porciani \etal breaks down when
the two points are highly correlated. In order to show this we
consider the limit in which $\nu_1=\nu_2=\nu$, $S_1=S_2=S$ and $\xi
\rightarrow S$, and restrict our attention to the physical range in
which $\delta_1, \delta_2 \leq \nu$. In this case our two-step
approximation simply reduces to the single absorbing barrier solution
as given by eq.\ (28),
\be
Q(\nu,\del_1,\del_2, S) =
[G(\del_1, S)-G(2 \nu -  \del_1, S)] \,
\del_D(\del_1 - \del_2)\ ,
\ee
where $G(\del,S)$ is a Gaussian of variance $S$ in the variable $\del$
[as in eq.\ (4)], and $\del_D$ is the one-dimensional Dirac delta
function. In this notation, the correlated solution with uncorrelated
boundary conditions as per eqs.\ (33) and (34) of Porciani \etal gives
instead
\be
Q(\nu,\del_1,\del_2, S) =
[G(\del_1, S) + G(2 \nu -  \del_1, S)] \,
\del_D(\del_1 - \del_2),
\ee
which amounts to an unphysical {\em increase}\/ of probability density
in the presence of absorbing barriers.  This is because as $\xi
\rightarrow S$, the double reflection as in the uncorrelated case
becomes an increasingly poor approximation to the single reflection
appropriate for correlated diffusion. Indeed, Figure 3 of Porciani
\etal shows that their approximation fails to reproduce a numerical
Monte Carlo solution when the two points are separated by a small
distance; their approximation for the clustering properties of dark
matter halos is accurate only for separations that are larger than the
Lagrangian halo size by a factor of $\sim 3$--10, with the exact value
depending on the power spectrum of density fluctuations and the halo
mass.

\section*{Appendix A: Evaluation of the Bivariate Mass Function}

In this Appendix we give various analytic expressions that are
necessary for the explicit evaluation of eq.\ (\ref{eq:littlef}). 
Using eq.\ (\ref{eq:Qpm}) to compute the derivatives, we find 
\ba
\frac{\partial^2 Q_\pm}
{\partial \del_1 \partial \del_2}
&=& \frac{1}{4 \pi (S_1 S_2 - \xi^2)^{3/2}}
\, \exp \left[ -\, \frac{\del_1^2 S_2 + \del_2^2 S_1 -
2 \del_1 \del_2 \xi} {2 (S_1 S_2 - \xi^2)}
\right]\ \times \nonumber \\
&& \left\{ \left[
	 \frac{(\del_1 S_2 - \del_2 \xi)
	        (\del_2 S_1  - \del_1 \xi)}
		{S_1 S_2 - \xi^2} + \xi \right]\,
\left[ {\rm erf} \left( \tilde \nu
\sqrt{\frac{\tilde S}{2}} \right) \pm 1 \right] \right.
\nonumber \\
&& -  \left. \sqrt{ \frac{2 \tilde S}{\pi} }
\exp \left(-\frac{\tilde S}{2} \tilde \nu^2 \right)
\left( \xi \frac{\nu_{\rm min}}{\tilde S} - \frac{\del_2 S_1}
	  {S_1 - \xi}
      - \frac{\del_1 S_2}
	{S_2 - \xi} \right) \right\}\ , \nonumber \\
\ea
and
\ba
\frac{\partial^2 Q_\pm}
{\partial \del_2^2}
&=& \frac{1}{4 \pi (S_1 S_2 - \xi^2)^{3/2}}
\, \exp \left[ -\, \frac{\del_1^2 S_2 + \del_2^2 S_1 -
2 \del_1 \del_2 \xi} {2 (S_1 S_2 - \xi^2)}
\right]\ \times \nonumber \\
&& \left\{ \left[
	 \frac{(\del_2 S_1  - \del_1 \xi)^2}
		{S_1 S_2 - \xi^2} - S_1 \right]\,
\left[ {\rm erf} \left( \tilde \nu
\sqrt{\frac{\tilde S}{2}} \right) \pm 1 \right]
\right.
\nonumber \\
&&
- \sqrt{ \frac{2 \tilde S}{\pi} }
\left.
\exp \left(-\frac{\tilde S}{2} \tilde \nu^2 \right)
\left[ \xi \frac{S_1-\xi}{S_2-\xi} \frac{\nu_{\rm min}}{\tilde S}
+ \frac{\del_1 \xi}{S_2-\xi} + \del_2 \frac{\xi^2+S_1 (\xi -2 S_2
)} {(S_2 - \xi)^2} \right] \right\}, \nonumber \\
\ea
while to compute  $\partial^2 Q_\pm / \partial \del_1^2$ we simply
switch the 1 and 2 indices in all terms of the previous equation.
Finally, we also require
\ba
\frac{\partial Q_\pm}
{\partial \xi}
&=& \frac{Q_\pm}{S_1 S_2-\xi^2} \left[\del_1 \del_2+\xi \left(1+\frac{
2\del_1\del_2\xi - \del_1^2 S_2 - \del_2^2 S_1} {S_1 S_2 -\xi^2} \right)
\right]\ + \nonumber \\
&& \frac{1}{4 \pi \sqrt{S_1 S_2 - \xi^2}} \, \exp \left[ -\,
\frac{\del_1^2 S_2 + \del_2^2 S_1 - 2 \del_1 \del_2 \xi} {2
(S_1 S_2 - \xi^2)} \right]\
\times \nonumber \\
&&
\exp \left(-\frac{\tilde S}{2} \tilde \nu^2 \right)
\sqrt{\frac{\tilde S}{2 \pi}} \left( 2 \frac{\partial \tilde \nu}
{\partial \xi} + \frac{\tilde \nu} {\tilde S} \frac{\partial \tilde
S} {\partial \xi} \right),
\ea
where \be \frac{\partial \tilde
S} {\partial \xi} = \frac{S_1 S_2 (S_2 - 2 \xi)(S_1-2 \xi)-\xi^4}
{(S_1 S_2 - \xi^2)^2}\ ,\ee and
\be \frac{\partial \tilde \nu} {\partial \xi} = -\frac{\nu_{\rm min}
} {\tilde S^2} \frac{\partial \tilde S} {\partial \xi} -\frac{\del_1}
{(S_1-\xi)^2}- \frac{\del_2} {(S_2-\xi)^2}\ , \ee although in our
standard cases this term is not needed as $\xi_k$ and $\xi_{r_{max}}$
are only functions of the smaller of any two given values $S_1$ and
$S_2$ [see eq.\ (\ref{eq:littlef})].

\section*{Appendix B: {\em Gemini},\,
	A Toolkit for Analytical Models of Inhomogeneous
	Structure Formation}

In order to facilitate applications of the analytical model of
inhomogeneous structure formation discussed in this paper, we have
made public a program {\em Gemini}\, that is intended to be a
toolkit that can be used to quickly evaluate the most important
quantities derived in this work. The program is divided into two main
sections; the first, {\em xiroutines},\, contains routines that
compute preliminary quantities including the mass correlation
function, in an arbitrary cosmology, and the second, {\em
gemini},\, uses these quantities to calculate the bivariate
functions described in this work.

The routines that calculate $\xi(r,d),$ $\frac{d \xi}{d
\sigma^2}(r,d),$ and $\sigma^2(r)$ need only be run once for any given
cosmology, as {\em xiroutines}\, constructs a table of such values
indexed as functions of $\ln(r)$ and $\ln(d-r)$ with arbitrary
resolution.  Subsequent calls to {\em xiroutines}\, then interpolate
between these values using a cubic spline technique.

The routines contained in {\em gemini},\, on the other hand, are
more fragmented, with the user determining which subroutines need to
be included into the main program in order to calculate the quantities
relevant to the problem at hand.  In this case no tabulation is used
and the bivariate mass function, $f$, is computed analytically. The
cumulative mass fraction and mixed mass function, on the other hand,
require a single numerical integration. These are carried out as a
function of the following four variables,
\be
s \equiv \frac{S_1-\xi}{\xi} \, ,\,
t \equiv \frac{\nu_1} {\sqrt{2 (S_1-\xi)}}\, , \,
u \equiv \frac{\nu_2 -\nu_1} {\nu_1}\,, \,
v \equiv  \sqrt{\frac{S_1-\xi}{S_2-\xi}}\, ,\,
\ee
which are chosen to have simple values in common limits such as $S_1 =
S_2$ or $\nu_1 = \nu_2.$ In terms of these quantities and in the case
in which $\nu_1 \leq \nu_2$
\be
F(\nu_1,\nu_2,S_1,S_2,\xi)=
\sqrt{\frac{s}{\pi}}\ J(s,t,u,v)\ ,
\label{eq:cum}
\ee
where \be
J(s,t,u,v)=\int_0^{\infty} dx\, \left[ e^{-s(t-x)^2}-
e^{-s(t+x)^2} \right]\,
{\rm erf}(x)\, {\rm erf}[v(t u+ x)]\ , \ee
while the case in which  $\nu_1 \leq \nu_2$, is calculated by simply switching
the 1 and 2 indices in eq.~(\ref{eq:cum}).

Finally, the $x$ integral for the mixed-mass function ${\cal F}$
can be written as
\be
\label{eq:mixedtable}
\frac{-1}{2 \pi \sqrt{\xi} (S_1-\xi)^{1/2}}\ \left[ (\nu_1-\nu_2)
\sqrt{\frac{2}{S_1-\xi}} \, K(s,t,u,v,w)+ 2 L(s,t,u,v,w)
\right]\ , \ee
where \be
K(s,t,u,v,w)=
\int_w^{\infty} dx\, \left[ e^{-s(t+t u-x)^2}-
e^{-s(t + t u -2 w+x)^2} \right]\, {\rm erf}(x v)\,
e^{-(x - t u)^2}\ , \ee
and \be
L(s,t,u,v,w)=\int_w^{\infty} dx\, \left[ e^{-s(t + t u -x)^2}-
e^{-s(t+ t u -2 w+x)^2} \right]\, x\, {\rm erf}(x v)\,
e^{-(x - t u)^2}\ . \ee
In this case the lower bound on the integration is defined as
$w \equiv \max(0 , t u).$

{\em Gemini}\, is available at the web address
{\tt http://www.arcetri.astro.it/$^{\sim}$evan/Gemini/} which also
contains more detailed documentation regarding its use.

\end{document}